\documentclass[11pt,twocolumn,preprintnumbers,amsmath,amssymb,nofootinbib,superscriptaddress,aps,prd,numberedappendix]{openjournal}
\usepackage{natbib}
\usepackage{newtxtext,newtxmath}
\usepackage{amsmath,amssymb}
\usepackage{graphicx}
\usepackage{dcolumn}
\usepackage{bm}
\usepackage[dvipsnames]{xcolor}
\usepackage{xspace}
\usepackage{ulem}
\usepackage[colorlinks,linkcolor=red,citecolor=blue,urlcolor=blue ]{hyperref}
\usepackage{graphicx}
\usepackage{multirow}
\usepackage{cancel}
\usepackage{booktabs}
\newcommand{\nv}{\hat{\bf n}}

\newcommand{\lcdm}{$\Lambda$CDM\xspace}

\newcommand{\nmt}{{\tt NaMaster}\xspace}
\newcommand{\quaia}{{\sl Quaia}\xspace}
\newcommand{\planck}{{\sl Planck}\xspace}
\newcommand{\gaia}{{\sl Gaia}\xspace}
\newcommand{\unWISE}{{\sl unWISE}\xspace}
\newcommand{\hpx}{{\tt HEALPix}\xspace}
\newcommand{\pantheonp}{Pantheon$+$\xspace}
\newcommand{\desyf}{DES-Y5\xspace}
\newcommand{\appropto}{\mathrel{\vcenter{
  \offinterlineskip\halign{\hfil$##$\cr
    \propto\cr\noalign{\kern2pt}\sim\cr\noalign{\kern-2pt}}}}}

\begin{document}
\title{Measurement of the power spectrum turnover scale from the cross-correlation between CMB lensing and {\sl Quaia}}

\author{David Alonso$^{1,*}$}
\author{Oleksandr Hetmantsev$^{2}$}
\author{Giulio Fabbian$^{3, 4}$}
\author{An\v ze Slosar$^{5}$}
\author{Kate Storey-Fisher$^{6}$}
\email{$^*$david.alonso@physics.ox.ac.uk}
\affiliation{$^1$Department of Physics, University of Oxford, Denys Wilkinson Building, Keble Road, Oxford OX1 3RH, United Kingdom}
\affiliation{$^2$Main Astronomical Observatory, National Academy of Sciences of Ukraine, 27 Akademik Zabolotnyi Street, Kyiv, 03143, Ukraine}
\affiliation{$^3$Kavli Institute for Cosmology Cambridge, Madingley Road, Cambridge CB3 0HA, UK}
\affiliation{$^4$Institute of Astronomy, Madingley Road, Cambridge CB3 0HA, UK}
\affiliation{$^5$Physics Department, Brookhaven National Laboratory, Upton NY11973, USA}
\affiliation{$^6$Kavli Institute for Particle Astrophysics and Cosmology, Stanford University, 452 Lomita Mall, Stanford, CA 94305, USA}
\date{\today}

\begin{abstract}
  We use the projected clustering of quasars in the \gaia-\unWISE quasar catalog, \quaia, and its cross-correlation with CMB lensing data from \planck, to measure the large-scale turnover of the matter power spectrum, associated with the size of the horizon at the epoch of matter-radiation equality. The turnover is detected with a significance of between $2.3$ and $3.1\sigma$, depending on the method used to quantify it. From this measurement, the equality scale is determined at the $\sim20\%$ level. Using the turnover scale as a standard ruler alone (suppressing information from the large-scale curvature of the power spectrum), in combination with  supernova data through an inverse distance ladder approach, we measure the current expansion rate to be $H_0=62.7\pm17.2\,{\rm km}\,{\rm s}^{-1}\,{\rm Mpc}^{-1}$. The addition of information coming from the power spectrum curvature approximately halves the standard ruler uncertainty. Our measurement in combination with calibrated supernovae from \pantheonp and SH0ES constrains the CMB temperature to be $T_{\rm CMB}=3.10^{+0.48}_{-0.36}\,{\rm K}$, independently of CMB data. Alternatively, assuming the value of $T_{\rm CMB}$ from COBE-FIRAS, we can constrain the effective number of relativistic species in the early Universe to be $N_{\rm eff}=3.0^{+5.8}_{-2.9}$.
\end{abstract}

\maketitle

\section{Introduction}\label{sec:intro}
  The study of the Universe's large-scale structure (LSS) aims to extract cosmological information from the distribution of matter inhomogeneities on different scales, and their evolution in time. Key to these studies is the linear matter power spectrum $P_L(k)$: the Fourier-space variance of the matter overdensity at early times\footnote{It is common to work with $P_L(k)$ at $z=0$ by simply evolving it using linear theory.}. The form of $P_L(k)$ is fully determined by a small number of characteristic scales, associated with specific physical phenomena affecting the growth of structure. On small scales, neutrino free streaming causes power suppression above a characteristic wavenumber $k_{\rm fs}$, related to the neutrino mass \citep{astro-ph/0603494,1109.4416}. Another key scale is the sound horizon of the baryon-photon plasma at the time of decoupling $r_d$, which sets the frequency of the baryon acoustic oscillations \citep[BAO,][]{astro-ph/9709112}, a central observable for LSS that brought the field into the regime of precision cosmology \citep{astro-ph/0501171}. Finally, the physical scale that gives rise to the most prominent feature in the matter power spectrum is the size of the horizon at the time of matter-radiation equality. Matter fluctuations on scales below this horizon were subject to a suppressed growth during the radiation-dominated era, whereas larger scales evolved largely unaffected into the matter epoch. This gives rise to a prominent feature in the matter power spectrum: a peak, or turnover, at a wavenumber close to the comoving horizon at the matter-radiation equality $k_{\rm eq}\sim0.01\,{\rm Mpc}^{-1}$, below which $P_L(k)$ grows as $\propto k^{n_s}$ (with $n_s$ the scalar spectral index), and above which it decays as $\sim k^{n_s-4}$. Specifically, this change of slope, from positive on large scales, to negative on small scales, and the scale at which it takes place, is what we will refer to as ``the turnover'' in what follows.

  The power spectrum turnover has significant value for cosmology. Much like the BAO scale, the turnover feature may be used as a standard ruler to place constraints on the distance-redshift relation. Furthermore, the dependence on cosmological parameters of the ruler (i.e. the size of the comoving horizon at matter-radiation equality) is highly complementary to that of other probes of the background expansion (e.g. supernovae and BAO), and thus can be used in combination with them to break parameter degeneracies and obtain complementary constraints on key parameters such as the late-time expansion rate $H_0$, the matter density parameter $\Omega_m$ and, potentially, the Dark Energy equation of state $w$. Interestingly, this may be achieved even without a clear detection of the turnover itself, relying instead on the broadband curvature in the power spectrum on smaller scales caused by the turnover. Several approaches recently employed in the literature, including full-shape power spectrum analyses \citep{2008.08084}, the {\sc ShapeFit} methodology \citep{2106.07641}, and exploiting the curvature of the power spectrum of the Cosmic Microwave Background (CMB) lensing convergence \citep{2204.02984}, benefit from this smaller-scale curvature.

  However, in spite of its prominence, the power spectrum turnover has been surprisingly difficult to detect directly in galaxy survey data. The main reason for this is the comparably large scale of this change in slope. Subtended on the sky, the equality scale would correspond to separations of about 10$^\circ$ at $z\sim0.5$ and 3$^\circ$ at $z=2$. Obtaining reliable measurements on such large scales from galaxy survey data is complicated for two reasons: first, samples covering very large volumes are required in order to reduce the statistical uncertainties sufficiently. Secondly, on these scales the observed galaxy overdensity can be dominated by observational systematics: the fluctuations in the observed number of galaxies caused by non-cosmological effects. These include Galactic contaminants, most prominently dust extinction and star contamination, as well as instrumental effects, such as variations in observing conditions leading to an inhomogeneous survey depth. In turn, measuring the small-scale slope of the power spectrum is significantly simpler. For this reason, to our knowledge, only two attempts have been made at detecting the power spectrum turnover from galaxy survey data. An early tentative detection was made with the WiggleZ survey \citep{1211.5605} and, more recently, the turnover was measured using quasars from the extended Baryon Oscillation Spectroscopic Survey, eBOSS \citep{2302.07484}. In both cases, these results were obtained using the 3D auto-correlation of galaxies, and thus significant effort had to be devoted to ensuring the robustness of the measurements against instrumental and survey systematics. Being able to make the same measurement using other LSS probes, with different sensitivity to these systematics, is therefore of vital importance.

  In this paper we present the first detection of the power spectrum turnover from the cross-correlation of the projected quasar overdensity and maps of the lensing convergence of the Cosmic Microwave Background (CMB), in combination with the projected clustering of these quasars. Since CMB lensing is insensitive to most of the sky systematics affecting galaxy number counts, the cross-correlation is particularly robust to the impact of large-scale contamination, and thus provides a powerful tool to study LSS on ultra-large scales \citep{1710.09465}. 
  We will make use of data from the \gaia-\unWISE quasar survey, \quaia, a full-sky quasar catalog covering one of the largest comoving volumes explored to date \citep{2306.17749}. Its reliance on space-based data provides additional robustness to sky contamination, and the availability of relatively precise spectro-photometric redshifts allows for a reliable tomographic analysis \citep[][]{2306.17748,2402.05761}. We will also explore the cosmological constraints enabled by this measurement in combination with external supernova data.

  The paper is structured as follows. Section \ref{sec:data} describes the datasets used in our analysis. Section \ref{sec:meth} describes the methodology employed, including the theoretical model underlying the expected signals, the data analysis techniques used, and the procedure employed to quantify the detectability of the turnover and measure its scale. Our results are then presented and discussed in Section \ref{sec:res}, and we conclude in Section \ref{sec:conc}.

\section{Data}\label{sec:data}
  \subsection{Quaia}\label{ssec:data.quaia}
    The \gaia-\unWISE quasar catalogue, \quaia, is a sample of $\sim1.3$ million quasars covering the full celestial sphere with magnitude $G<20.5$. The catalogue was constructed by matching the sample of \gaia quasar candidates with \unWISE sources to improve the sample purity and the quality of the source redshifts. The spectro-photometric redshifts $z_Q$ were estimated from the combined \gaia spectra and \unWISE photometry via $k$-nearest neighbour matching against the Sloan Digital Sky Survey DR16Q quasar sample.

    We divide the full sample, spanning the redshift range $0< z \lesssim 4$, into the same 2 redshift bins used in \cite{2306.17748}, corresponding to sources with redshifts above and below the median redshift of the sample $z_Q^{\rm med}=1.47$ (we will refer to these as the ``High-$z$'' and ``Low-$z$'' bins, respectively). The ability to select different redshift ranges is extremely useful in this case, as it allows us to quantify whether the projected turnover scale scales with redshift as expected. We will also study the results obtained from the full sample as an additional internal consistency test. A selection function $w(\nv)$ was generated for each redshift bin characterising the fluctuations in the mean density of sources caused by spatially-varying systematics. This selection function is generated via Gaussian process reconstruction as described in \cite{2306.17749}. As described in Section \ref{sssec:meth.cls.cls}, we deproject a number of systematic templates from the quasar ovedensity maps prior to power spectrum estimation. These include the templates used to create the selection functions.

    From the catalog and selection function, galaxy overdensity maps are generated as
    \begin{equation}
      \delta_g(\nv)=\frac{N(\bf n)}{\bar{N}w(\nv)}-1,
    \end{equation}
    where $N(\nv)$ is the number of quasars found in the pixel lying along the direction $\nv$, and the mean number of quasars per pixel is
    \begin{equation}
      \bar{N}\equiv\frac{\sum_{\nv}N(\nv)}{\sum_{\nv}w(\nv)}.
    \end{equation}
    As in \cite{2306.17748,2402.05761}, we mask out all pixels where the selection function is $w<0.5$. All maps were generated using the \hpx pixelisation scheme with resolution parameter $N_{\rm side}=256$, corresponding to $0.22^\circ$-sized pixels. The redshift distribution of each bin is estimated by stacking the individual redshift probability density functions of all quasars in the bin, which are parametrised as normal distributions. This was determined to be a sufficiently accurate procedure in \cite{2306.17748}.
    
  \subsection{Planck PR4 lensing map}\label{ssec:data.kappa}
    We use CMB lensing maps reconstructed from the fourth \planck public data release (PR4), generated through the {\tt NPIPE} pipeline \citep{2007.04997}, and presented in \cite{2206.07773}. In particular, we made use of the ``Generalised Minimum Variance'' (GMV) convergence map, which implements a joint inverse-variance Wiener filtering of the temperature and polarisation data accounting for inhomogeneous noise, achieving a $\sim20\%$ better signal-to-noise ratio than previous releases. Although the cosmological analysis of the CMB lensing auto-spectrum used the multipole range $\ell\geq 8$ \citep{2206.07773}, we compute the cross-correlation with \quaia down to $\ell=2$. Our fiducial analysis will use scales $\ell\geq 4$, and we will show that the results are insensitive to this choice. The choice $\ell\geq 8$ for the \planck lensing power spectrum analysis stems from concerns about the reliability of the subtracted mean field \cite{1807.06210}, which is less relevant for cross-correlations both with external data \citep{2305.07650} and other Planck CMB observables \citep{Carron2022}.

    \cite{2306.17748} identified the potential presence of residual systematics in the CMB lensing map when cross-correlated with high-redshift \quaia sources. A more detailed study by \cite{2402.05761} showed that the evidence for these is inconclusive, and that if present, this contamination does not impact cosmological analyses significantly. The resulting redshift bins are centered at $\bar{z}=1.0$ and 2.1 (mean redshift).

    The GMV map is provided in terms of its harmonic coefficients $\kappa_{\ell m}$. We generate the CMB lensing map at the desired $N_{\rm side}=256$ by masking out all harmonic coefficients with $(\ell,m)\geq3\times N_{\rm side}$, and applying an inverse spherical harmonic transform to the resulting coefficients. We use the \planck CMB lensing analysis mask described in \cite{2206.07773}.

\section{Methods}\label{sec:meth}
  \subsection{Power spectra and covariances}\label{ssec:meth.cls}
    \subsubsection{Power spectrum estimation}\label{sssec:meth.cls.cls}
      We estimate all power spectra through the pseudo-$C_\ell$ or ``MASTER'' algorithm \citep{astro-ph/0105302}, as implemented in \nmt\footnote{\url{https://github.com/LSSTDESC/NaMaster}} \citep{1809.09603}. The pseudo-$C_\ell$ algorithm is a particular implementation of the more general family of quadratic power spectrum estimators, which effectively assumes that the covariance matrix of the target maps is diagonal \citep{astro-ph/9611174,astro-ph/0307515,1306.0005}. This allows one to use fast analytical methods to characterise the impact of mode-coupling due to the presence of a sky mask. The downside of pseudo-$C_\ell$ algorithm is its sub-optimality, particularly for signal-dominated data with a steep power spectrum on large scales (where the approximation of a diagonal covariance matrix is not valid). In our case, however, given the relatively high level of shot noise, this approximation is rather good, even on large scales. Repeating our calculation using a minimum-variance version of the quadratic estimator\footnote{Note that the implementation of the quadratic estimator used for this test did not incorporate the transfer function correction for systematic deprojection we discuss in Section \ref{sssec:meth.cls.tf}. As we show in Section \ref{sec:res}, the impact of this transfer function on our results is nevertheless negligible.} \citep{astro-ph/9611174}, we verified that the pseudo-$C_\ell$ approach is able to recover equivalent power spectrum uncertainties throughout the whole range of scales explored here.

      We measure all power spectra in bandpowers ($\ell$ bins) of width $\Delta \ell=2$ in the range $0\leq\ell<40$, $\Delta \ell=5$ in the range $40\leq\ell<60$, and $\Delta \ell=30$ in the range $\ell\geq60$. This binning allows us to capture the large-scale behaviour of the power spectrum with sufficient resolution, while limiting the total number of elements in the data vector to a reasonable value. 

      Although the space-based nature of the data used to construct \quaia allows us to significantly reduce the impact of large-scale systematic contamination in the sample, the quasar auto-correlation displays a clear excess at very large scales ($\ell\lesssim10$ -- see Fig. \ref{fig:cl_master}). Although these additive systematics do not bias the cross-correlation with CMB lensing (assuming that the CMB lensing map is free from systematics that correlate with those presence in the galaxy map), their presence leads to an increase in the variance of the estimated power spectrum on large scales, which degrades our ability to constrain the turnover scale. To mitigate this, we use template deprojection \citep{1609.03577} to remove any linear trend in the quasar overdensity maps proportional to a set of templates that represent the spatial variation of relevant systematic diagnostics. This set includes: (i) a Galactic dust extinction template from \cite{2306.03926}, (ii) two star templates constructed by randomly selecting sources from \gaia and \unWISE, respectively, (iii) templates tracking the scanning patterns of \gaia and \unWISE, and (iv) similar star templates targetting only the region around the large and small Magellanic clouds (see \cite{2306.17749} for further details). In addition to these, we further remove the dipole component of the overdensity maps by deprojecting three templates corresponding to the $\ell=1$ spherical harmonic functions (specifically, $Y_{10}(\nv)$, ${\rm Re}(Y_{11}(\nv))$, and ${\rm Im}(Y_{11}(\nv))$) \citep{WilliamsInprep}.
      This removes a large dipole component at the map level that reduces the power spectrum variance at $\ell\geq2$ due to mode coupling. We remove the mean across the survey mask of all systematic templates except for the dipole templates.

    \subsubsection{Transfer functions}\label{sssec:meth.cls.tf}
      \begin{figure*}
        \centering
        \includegraphics[width=0.95\textwidth]{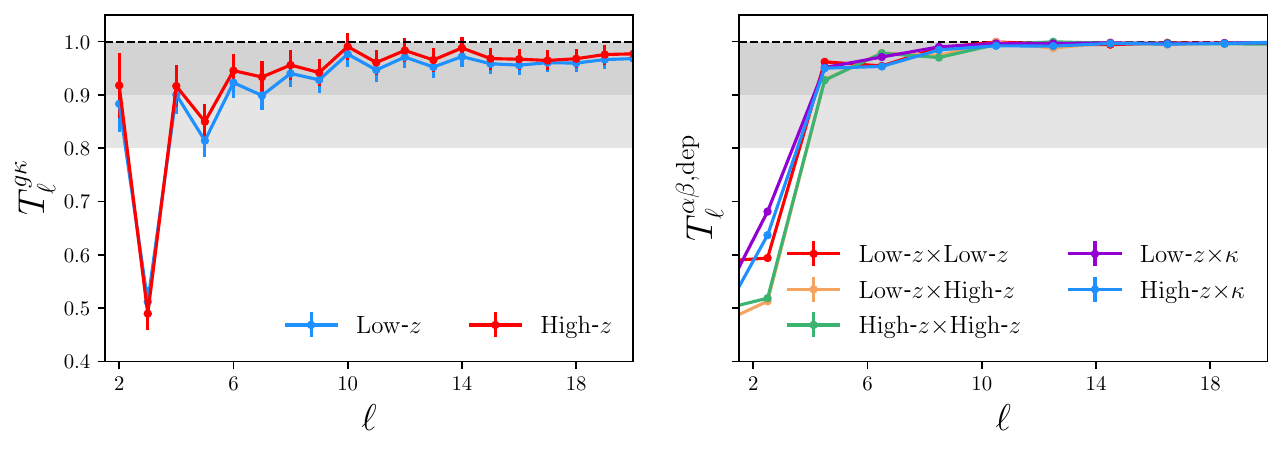}
        \caption{Lensing (left) and deprojection (right) transfer functions for the different relevant cross-correlations.}
        \label{fig:transfer}
      \end{figure*}
      Since mode-deprojection removes all power from the map in modes proportional to any of the systematic templates deprojected, it leads to a bias in the estimated power spectra due to mode loss. This bias usually negligible on the relatively small scales used for cosmological analyses unless a very large number of templates is deprojected. However, on the large scales explored here, this bias must be characterised and corrected. As described in \cite{1609.03577,1809.01669},  this can be done analytically assuming an accurate knowledge of the true underlying power spectrum. Since this may not be possible on very large scales, where measurement uncertainties prevent us from obtaining precise constraints on the power spectrum, we instead follow an alternative, simulation-based approach. Consider the problem of characterising the impact of mode deprojection in the power spectrum between two maps, $m_1$ and $m_2$, with masks $w_1$ and $w_2$, respectively. To do so, we generate a Gaussian random map $s$ following a given input power spectrum\footnote{We used $C_\ell\propto1/(1+\ell/\ell_*)$ with $\ell_*=30$, but verified that the result of this calculation was insensitive to the choice of input power spectrum.}. We then generate two masked versions of this map, $s_1$ and $s_2$, for the two masks involved, and deprojected versions of the same maps, $\tilde{s}_1$ and $\tilde{s}_2$ (note that, if one of the fields involved is the CMB lensing convergence map, no deprojection is applied to the corresponding simulated map). We then calculate an effective transfer function for the power spectrum as
      \begin{equation}
        T^{12,{\rm dep}}_\ell\equiv\frac{\langle C_\ell(\tilde{s}_1,\tilde{s}_2)\rangle_s}{\langle C_\ell(s_1,s_2)\rangle_s},
      \end{equation}
      where $C_\ell(a,b)$ is the power spectrum between maps $a$ and $b$, and $\langle\cdots\rangle_s$ denotes averaging over Gaussian simulations. The unbiased power spectrum for maps $m_1$ and $m_2$ is then given by
      \begin{equation}
        C_\ell(m_1,m_2)=\frac{\tilde{C}_\ell(m_1,m_2)}{T^{12,{\rm dep}}_\ell},
      \end{equation}
      where $\tilde{C}_\ell$ is the power spectrum calculated from mode-deprojected maps without correcting for the impact of this deprojection.

      In a similar manner, we must account for the effects of mode loss in the lensing reconstruction maps. The presence of a sky mask leads to a mis-normalisation of the reconstructed lensing map. Since this effect is inhomogeneous, its impact on the final power spectrum depends on the sky mask of the fields being cross-correlated \citep{1301.4145,1807.06210,2210.05449,2305.07650}. We estimate the transfer function associated with this effect making use of the lensing reconstruction simulations made available with the PR4 lensing maps\footnote{See \url{https://github.com/carronj/planck_PR4_lensing}.}, as described in \cite{2309.05659}. For each realisation, three maps are generated: $\kappa_\kappa$, the input convergence field masked with the CMB lensing analysis mask, $\kappa_g$, the input convergence field masked with the \quaia selection function, and $\kappa_{\rm rec}$, the reconstructed convergence field (masked with the lensing mask). The transfer function is then:
      \begin{equation}
        T^{g\kappa}=\frac{\langle C_\ell(\kappa_{\rm rec},\kappa_g)\rangle_\kappa}{\langle C_\ell(\kappa_\kappa,\kappa_g)\rangle_\kappa},
      \end{equation}
      where $\langle\cdots\rangle_\kappa$ represents averaging over all $\kappa$ simulations.

      Figure \ref{fig:transfer} shows the transfer function due to deprojection for the 5 different power spectra considered here (auto- and cross-correlations between the two quasar redshift bins and their cross-correlations with $\kappa$), and the lensing reconstruction transfer function for the two lensing cross-correlations. In both cases, the transfer function is large $T_\ell\sim0.5$ at $\ell\leq3$, rising rapidly to $T_\ell\gtrsim0.8$-0.9 at higher $\ell$. As we will see, given our scale cuts, this effect is largely absorbed by the amplitude parameters of the model, and the impact of these transfer functions on our final results is small. Note that our fiducial scale cut is $\ell\geq4$ for the CMB lensing cross-correlations, and $\ell\geq15$ for the quasar auto-correlations.

    \subsubsection{Covariance matrix}\label{sssec:meth.cls.cov}
      \begin{figure}
          \centering
          \includegraphics[width=0.5\textwidth]{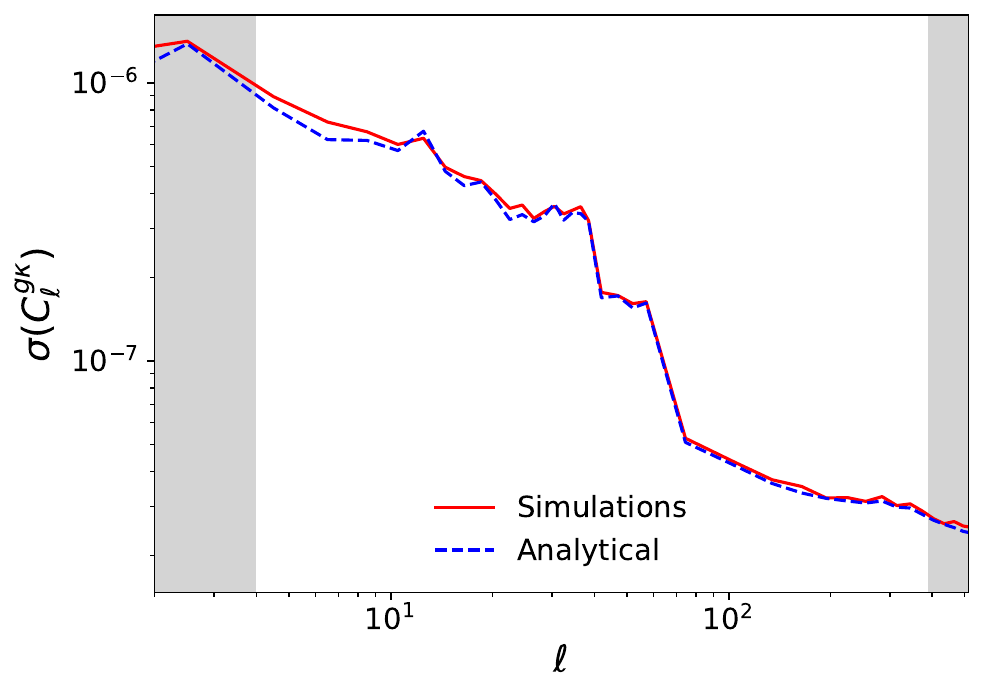}
          \caption{Power spectrum uncertainties estimated analytically (dashed blue) and via Gaussian simulations (solid red). The approximate analytical approach is in reasonable agreement with the simulation-based covariance used in our analysis, with the largest difference between them at the level of $\sim15\%$.}
          \label{fig:cov_comp}
      \end{figure}
      Accurate analytical methods have been developed for the calculation of angular power spectrum covariance matrices, accounting for the effects of mode coupling and the LSS-induced non-Gaussianity \citep{astro-ph/0307515,1807.04266,1906.11765}. Unfortunately, often the approximations involved lose accuracy on the largest scales, which are particularly important in our analysis. For this reason, we estimate the covariance matrix of our measurements through an empirical, simulation-based approach.

      We generate 1000 Gaussian signal-only simulations of the quasar overdensity and CMB convergence maps, and measure their power spectra following the same steps taken in the analysis of the real data (masking, deprojection, transfer function corrections). The sample covariance matrix is then estimated from the measured spectra of these simulations. We use the measured power spectra of the data as input to generate these simulations, thus ensuring that the covariance incorporates the additional variance caused by the presence of residual systematics in the \quaia maps. This procedure neglects a number of potentially important effects:
      \begin{itemize}
        \item {\bf Noise inhomogeneity.} Both \quaia and the \planck $\kappa$ maps contain inhomogeneous noise, and its contribution to the covariance matrix is not captured by statistically homogeneous simulations characterised by a set of power spectra. We verified that the impact of inhomogeneity caused by depth fluctuations in \quaia on the power spectrum uncertainty is negligible (percent-level variations) by generating Poisson realisations following the \quaia selection function. We verified that the impact of noise inhomogeneity in the $\kappa$ map is small by using the official \planck lensing simulations.
        \item {\bf Noisy input power spectra.} Statistical fluctuations in the measured power spectra used to generate the Gaussian simulations could lead to significant over- or under-estimation of the statistical uncertainties. We verified that replacing the measured power spectra by low-order polynomial fits to these spectra (in logarithmic space), did not affect the results presented in Section \ref{sec:res} in terms of detection significance of the power spectrum turnover or precision of the turnover scale measurement. 
      \end{itemize}
      The procedure outlined above only captures the purely disconnected power spectrum covariance, missing any connected contributions caused by non-Gaussianity in the galaxy distribution. This was determined to be a reasonable approximation in \cite{2306.17748}, and is a less important problem for the large scales used in this analysis.
      
      As an additional final test for the validity of our estimated covariance, we compare it against the analytical prediction generated by \nmt, described in \cite{1906.11765} (the so-called narrow-kernel approximation (NKA)). Fig. \ref{fig:cov_comp} shows the result of this comparison for the cross-correlation between \planck and the second \quaia redshift bin (which dominates the constraints presented in Section \ref{sec:res}). The figure shows the statistical uncertainties in the measured cross-spectrum uncertainties calculated via simulations (red solid line) and analytically (blue dashed line). The analytical prediction matches the simulated covariance reasonably well, with a maximum deviation between both error estimates of $\sim15\%$ at $\ell\sim7$.

  \subsection{Theory model and likelihood}\label{ssec:meth.th}
    \subsubsection{Angular power spectra}\label{sssec:meth.th.cl}
      \begin{figure}
        \centering
        \includegraphics[width=0.48\textwidth]{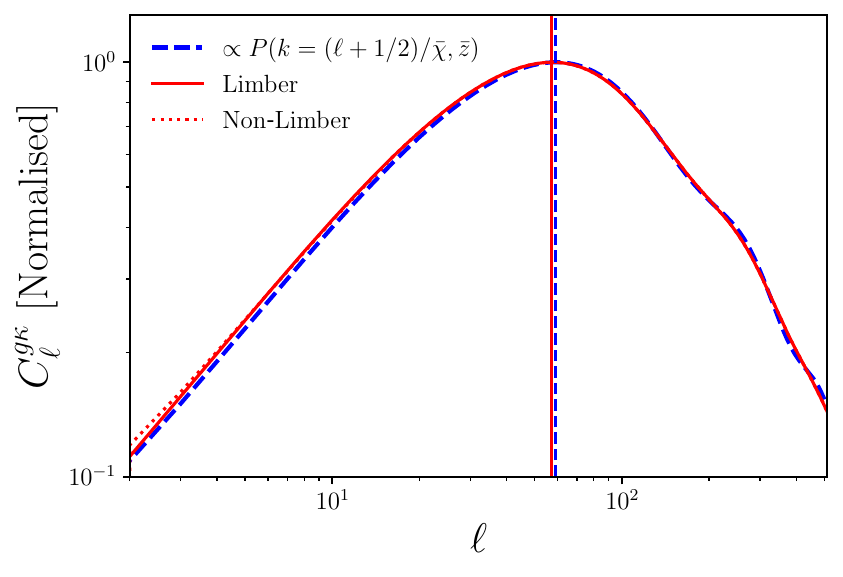}
        \caption{Theoretical prediction for the angular cross-spectrum between the second \quaia redshift bin and CMB lensing using Limber's approximation (solid red), and with no approximations (dotted red). The dashed blue line shows the 3D matter power spectrum evaluated at the scales suggested by the Limber projection $k=(\ell+1/2)/\bar{\chi}$, where $\bar{\chi}$ is the comoving distance to the median redshift of the sample. All curves are scaled by a multiplicative factor to ensure they have same amplitude. The two vertical lines mark the position of the maximum of the red and blue curves (i.e. the projected turnover scale).}\label{fig:cl_limber}
      \end{figure}
      Our main observable is the angular cross-power spectrum between the projected overdensity of sources in \quaia, $\delta_g(\nv)$, and the CMB lensing convergence map $\kappa(\nv)$ reconstructed from \planck data. Both projected quantities are related to the three-dimensional quasar and matter overdensities, $\Delta_g({\bf x},z)$ and $\Delta_m({\bf x},z)$, respectively, via
      \begin{align}
        &\delta_g(\nv)=\int d\chi\,q_g(\chi)\Delta_g(\chi\nv,z(\chi)),\\
        &\kappa(\nv)=\int d\chi\,q_\kappa(\chi)\Delta_m(\chi\nv,z(\chi)),
      \end{align}
      where $\chi$ is the radial comoving distance, and the radial kernels $q_g$ and $q_\kappa$ are
      \begin{equation}
        q_g(\chi(z))\equiv H(z)\,p(z),\hspace{12pt}
        q_\kappa(\chi)\equiv\frac{3H_0^2\Omega_m\chi}{a(\chi)}\frac{\chi_*-\chi}{\chi_*},
      \end{equation}
      where $p(z)$ is the redshift distribution of the galaxy sample under study, $H(z)$ is the expansion rate at redshift $z$, $H_0$ is its value today, $\Omega_m$ is the fractional energy density in non-relativistic species today, and $\chi_*$ is the distance to the last-scattering surface. Note that we will assume a flat $\Lambda$CDM cosmology in our analysis, in which case the radial comoving distance $\chi(z)$ is equivalent to the comoving angular diameter distance $d_A(z)$, which we will refer to in Section \ref{sssec:meth.th.cosmo}.

      The statistics of two projected fields $u$ and $v$ (which stand for any of $\delta_g$ or $\kappa$) are related to their 3D counterparts $U$ and $V$ (i.e. $\Delta_g$ or $\Delta_m$) and, specifically, their angular and 3D power spectra are related via
      \begin{align}\nonumber
        C_\ell^{uv}=\frac{2}{\pi}\int &dk\,d\chi_1\,d\chi_2\,j_\ell(k\chi_1)\,\,j_\ell(k\chi_2)\\ \label{eq:cl_nonlim}
        &q_u(\chi_1)\,q_v(\chi_2)\,P_{UV}(k,z_1,z_2),
      \end{align}
      where $j_\ell(x)$ is the spherical Bessel function of order $\ell$. We will approximate all unequal time power spectra as
      \begin{equation}
        P_{UV}(k,z_1,z_2)\simeq \sqrt{P_{UV}(k,z_1)P_{UV}(k,z_2)}.
      \end{equation}
      For sufficiently wide radial kernels, the expression above can be simplified significantly by approximating the Bessel functions as Dirac delta functions in the so-called Limber approximation \citep{1953ApJ...117..134L}. The result is
      \begin{equation}\label{eq:cl_lim}
        C_\ell^{uv}=\int \frac{d\chi}{\chi^2}\,q_u(\chi)q_v(\chi)\,P_{UV}(k=(\ell+1/2)/\chi,z(\chi)).
      \end{equation}
      Since Limber's approximation loses accuracy on large scales, which are critical for our goal of constraining the large-scale power spectrum turnover, we will not use it in our analysis, and instead perform the full 3D integral in Eq. \ref{eq:cl_nonlim}. In particular we make use of the {\tt FKEM} method presented in \cite{1911.11947}, as implemented in the Core Cosmology Library {\tt CCL} \citep{1812.05995} (see \cite{2212.04291} for details). 
      
      Nevertheless, Limber's approximation is useful to illustrate what we should expect when constraining the 3D power spectrum turnover from projected statistics. If any of the radial kernels in Eq. \ref{eq:cl_lim} is relatively narrow, with support only around a mean distance $\bar{\chi}$, it is easy to see that the resulting angular power spectrum will preserve the scale dependence of the 3D power spectrum, with a relatively well-defined mapping between angular and physical scales:
      \begin{equation}\label{eq:cl_approx}
        C_\ell \appropto P(k=(\ell+1/2)/\bar{\chi},\bar{z}).
      \end{equation}
      Thus, assuming the galaxy samples used have a sufficiently narrow redshift distribution\footnote{Note that, although the redshift bins used here are not particularly narrow, Fig. \ref{fig:cl_limber} shows that this approximation is still remarkably accurate.}, a turnover in $P(k)$ should be detectable as a similar turnover in $C_\ell$ at an angular scale that grows with the mean redshift of the sample under study. 
      \begin{figure*}
          \centering
          \includegraphics[width=0.9\textwidth]{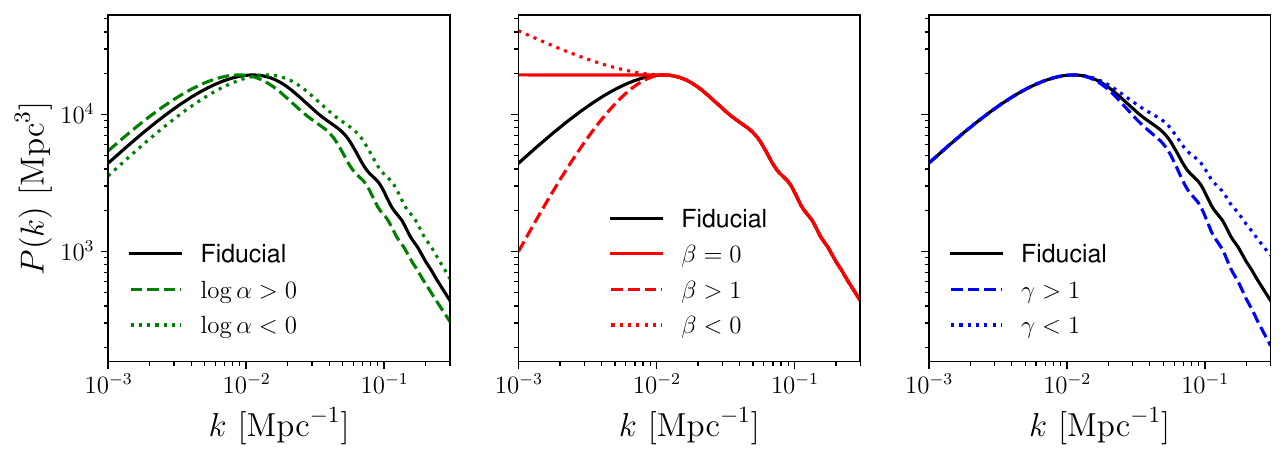}
          \caption{Effect in the power spectrum parametrisation of changing the stretch parameter $\alpha$ (left), the large-scale slope $\beta$ (centre), and the small-scale slope $\gamma$ (right).}
          \label{fig:pk_params}
      \end{figure*}
      
      This is illustrated in Fig. \ref{fig:cl_limber}. The figure shows the angular cross-power spectrum between quasars and CMB lensing predicted for the second \quaia redshift bin. The predictions using the  Limber approximation and with no approximations are shown as solid and dotted red lines respectively. The dashed blue line shows the approximation in Eq. \ref{eq:cl_approx}, in which the 3D power spectrum is simply evaluated at the projected scale $k=(\ell+1/2)/\bar{\chi}$, and scaled by an arbitrary amplitude. The main impact of the radial projection along the broad radial kernel of the \quaia sample is a damping of the BAO wiggles (a known effect, see \cite{1006.3226}) and a small change in the large-scale slope of the spectrum. The angular power spectrum, however, peaks very close to the expected projected turnover scale, which is therefore highly robust against projection effects (the position of the maxima for $C_\ell$ and $P(k)$ is marked by the vertical lines). The figure also shows that non-Limber corrections are fairly small on all scales used here (even though we will not apply Limber's approximation in our analysis).
      
    \subsubsection{Modelling the power spectrum turnover}\label{sssec:meth.th.model}
      Assuming a linear bias model, valid on the large scales used here (see Section \ref{sssec:meth.th.like}), the galaxy-galaxy and galaxy-matter power spectrum are simply proportional to the matter power spectrum
      \begin{equation}
        P_{gg}(k,z)=b_g(z)^2P_{mm}(k,z),\hspace{6pt} P_{gm}(k,z)=b_g(z)P_{mm}(k,z),
      \end{equation}
      where $b_g(z)$ is the redshift-dependent linear Galaxy bias.

      Our main aims in this analysis are:
      \begin{enumerate}
        \item Detecting the turnover of the matter power spectrum at large scales, and quantifying the confidence level of this detection.
        \item Determining the scale at which this turnover takes place.
      \end{enumerate}
      To address these goals, we parametrise the matter power spectrum as
      \begin{equation}
        P_{mm}(k)=A\,P_{\rm fid}(\alpha k)\,\left(\frac{P^{\rm LS}_{\rm flat}(\alpha k)}{P_{\rm fid}(\alpha k)}\right)^{1-\beta}\left(\frac{P^{\rm SS}_{\rm flat}(\alpha k)}{P_{\rm fid}(\alpha k)}\right)^{1-\gamma},
      \end{equation}
      where we have omitted the redshift dependence of the power spectrum for simplicity. $P_{\rm fid}(k)$ is the matter power spectrum in a fiducial cosmology. We calculate $P_{\rm fid}(k)$ using {\tt CAMB}\footnote{\url{https://camb.readthedocs.io}} \citep{astro-ph/9911177} to compute the linear matter power spectrum for the best-fit cosmological parameters measured by \planck \citep{1807.06209}, and then calculate the corresponding non-linear power spectrum using the {\tt HALOFIT} prescription \citep{1208.2701}. As we show in Section \ref{sec:res}, the specific choice of power spectrum template used does not impact our results significantly. $P^{\rm LS}_{\rm flat}(k)$ and $P^{\rm SS}_{\rm flat}(k)$ are defined to be flat on scales larger and smaller, respectively, than the turnover scale in the fiducial model:
      \begin{align}
        P_{\rm flat}^{\rm LS}(k)=\left\{
        \begin{array}{cc}
          P_{\rm fid}(k_{\rm TO}) & k < k_{\rm TO} \\
          P_{\rm fid}(k) & k \geq k_{\rm TO}
        \end{array}
        \right.,\\
        P_{\rm flat}^{\rm SS}(k)=\left\{
        \begin{array}{cc}
          P_{\rm fid}(k) & k < k_{\rm TO} \\
          P_{\rm fid}(k_{\rm TO}) & k \geq k_{\rm TO}
        \end{array}
        \right..
      \end{align}
      $k_{\rm TO}$ is the turnover scale, which we find by maximising $P_{\rm fid}(k)$.

      This model has a total of 5 free parameters, including the galaxy bias $b_g$ and:
      \begin{itemize}
        \item The {\bf prefactor} $A$ parametrises any potential deviations in the amplitude of the true power spectrum with respect to the fiducial model.
        \item The {\bf large-scale slope} $\beta$ controls the tilt of the power spectrum on large scales. A value of $\beta=0$ would correspond to a power spectrum that reaches a plateau instead of a clear turnover, with negative values corresponding to monotonically decreasing power spectra. $\beta=1$ recovers the fiducial large-scale behaviour, while $\beta>1$ corresponds to a model with an even sharper turnover.
        \item The {\bf stretch} parameter $\alpha$ controls the scales at which features in the power spectrum (most prominently the turnover) appear, with $\alpha=1$ corresponding to the fiducial cosmology.
        \item The {\bf small-scale slope} $\gamma$ parametrises the tilt of the power spectrum on small scales, and allows us to ensure that any constraints obtained on $\alpha$ are not driven by the small-scale curvature of the power spectrum, but by the location of the turnover feature.
      \end{itemize}
      The effects of varying $\alpha$, $\beta$, and $\gamma$ on the power spectrum are shown in Fig. \ref{fig:pk_params}.
      
      To simplify the theory model and associated likelihood, described in the next section, we will combine the bias and prefactor parameters into two pure amplitude parameters, $A_{gg}\equiv b_g^2 A$ and $A_{gm}\equiv b_gA$, multiplying the galaxy-galaxy and galaxy-matter power spectra respectively. Our full set of free parameters is therefore $\{A_{gg},A_{gm},\alpha,\beta,\gamma\}$. From the constraints on these parameters we will address the two aims listed above: the detection significance of the power spectrum turnover will be determined from the preference of the data for positive values of $\beta$, and the position of this turnover will be determined by the measured value of $\alpha$.
      \begin{figure*}
          \centering
          \includegraphics[width=0.9\textwidth]{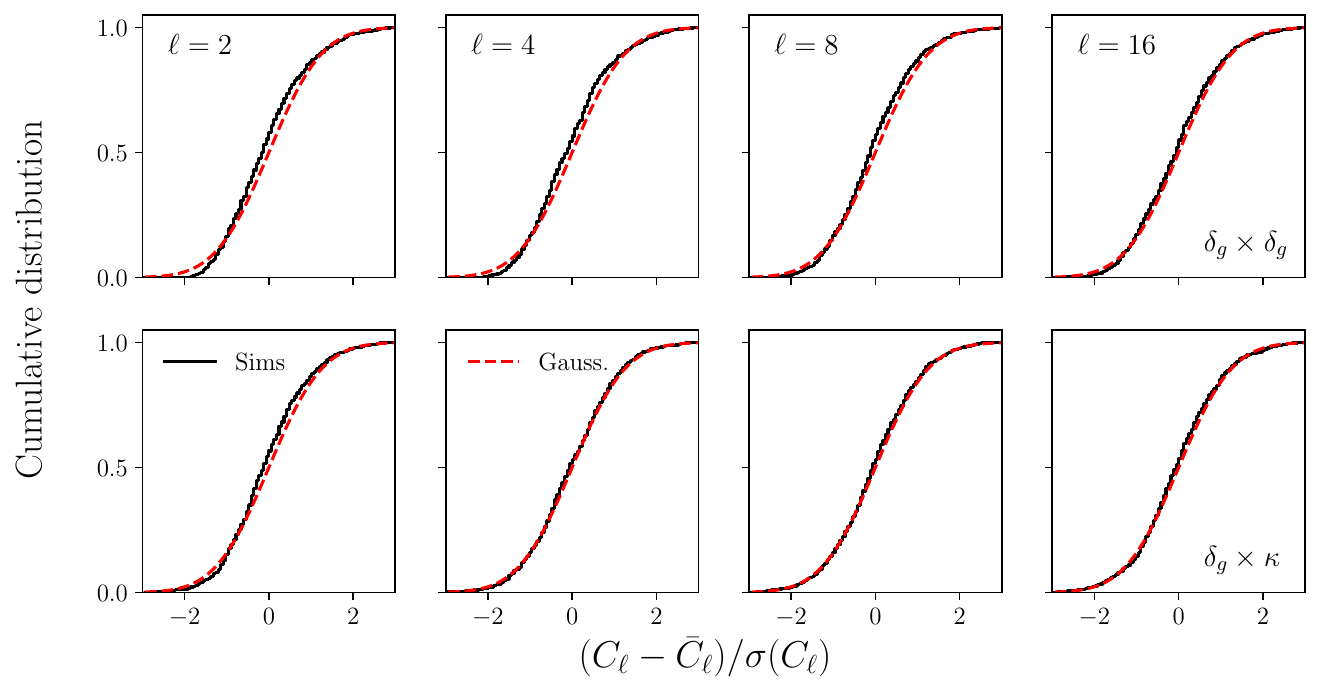}
          \caption{Cumulative distribution function of low-$\ell$ power spectra. Results are shown for the quasar auto-correlation in the High-$z$ redshift bin (top row) and for its cross-correlation with $\kappa$ (bottom row). The distribution from map-level simulations is shown in black, with the Gaussian approximation in dashed red. On the scales used in our fiducial analysis ($\ell\geq4$ for $\delta_g\times\kappa$ and $\ell\geq16$ for $\delta_g\times\delta_g$), the Gaussian approximation reproduces the true distribution reasonably well, while departures from it are clearly noticeable for the low-$\ell$ autocorrelation.}
          \label{fig:cl_dist}
      \end{figure*}

      Finally, we note that we assume the same redshift dependence for the quasar bias used in \cite{2306.17748,2402.05761}, based on the measurements of \cite{1705.04718} (with the overall normalisation of the $b_g(z)$ function absorbed by the $A_{gg}$ and $A_{gm}$ amplitude parameters). To further absorb any deviation from this particular redshift dependence in our parametrisation, we include two separate pairs of amplitude parameters, one for each redshift bin. We further verified that using the fitting function found by \cite{2402.05761} for the \quaia sample did not change the results of our analysis.

    \subsubsection{Likelihood}\label{sssec:meth.th.like}
      To constrain the free parameters of this model we use a data vector consisting of measurements of the \quaia auto-correlation and its cross-correlation with $\kappa$ in the two redshift bins described in Section \ref{ssec:data.quaia}.
      
      The presence of residual systematic contamination in the quasar auto-spectrum prevents us from using it on large scales. We thus impose a fiducial large-scale cut on the auto-correlations of $\ell_{\rm min}^{gg}=15$. This choice of scale was determined by quantifying the the difference between the power spectra computed with and without contaminant deprojection and removing scales on which these differences could not be explained by statistical fluctuations. The cross-correlation is significantly more robust to systematic contamination, and therefore we use $\ell_{\rm min}^{g\kappa}=4$ as a fiducial large-scale cut. As shown in Fig. \ref{fig:transfer}, the modes with $\ell<4$ are significantly more affected by mode loss due to deprojection and lensing reconstruction. This could lead to biases due to inaccuracies in the estimated transfer function. In practice, we will show that the constraints presented here are relatively insensitive to these large-scale cut choices.
      
      To ensure that the linear bias model used here is valid, we discard angular multipoles above $\ell_{\rm max}=k_{\rm max}\bar{\chi}$, where $k_{\rm max}=0.07\,{\rm Mpc}^{-1}$, and $\bar{\chi}$ is the comoving distance to the mean redshift of the sample. This corresponds to $\ell_{\rm max}=236$ and 390 in the low- and high-redshift bins respectively. With the power spectrum binning described in Section \ref{sssec:meth.cls.cls}, this results in a total of $N_d=110$ power spectrum measurements in our fiducial analysis (combining $C_\ell^{gg}$ and $C_\ell^{g\kappa}$ for the two \quaia redshift bins).

      We approximate the likelihood of this data vector as a multivariate normal distribution. Although this is often a bad approximation for power spectra measured on large scales, where the central limit theorem (CLT) does not apply, we find it is actually applicable in our case. The reason is that, due to large-scale contamination, the galaxy auto-correlation measurements are only used on relatively small scales, where the CLT can be applied. In turn, although we do use the cross-spectrum down to $\ell=4$ (and even $\ell=2$), the likelihood for cross-spectra is significantly more Gaussian than for auto-spectra (see e.g. \cite{2305.07650,2406.19488}). To validate this approach, we study the distribution of power spectrum measurements from the set of 1000 simulations used to compute the covariance matrix (described in Section \ref{sssec:meth.cls.cov}). The cumulative distribution function for low-$\ell$ bandpowers in both auto- and cross-correlations are shown in Fig. \ref{fig:cl_dist}. Although clear departures from a normal distribution are visible at the lowest $\ell$s for the quasar auto-correlation, the Gaussian approximation is able to reproduce the true distribution reasonably well on the scales used in our fiducial analysis. We quantify this by obtaining best-fit values for the model parameters in the 1000 simulations described above, and in 1000 simulated sets of power spectra drawn from a multivariate normal distribution. We then compare the distribution of the best-fit parameters obtained from both simulations. The mean values of $\alpha$ recovered in both cases differ by less than 2\% of the statistical uncertainties, while the standard deviation of the distributions agree at the $4\%$ level. A Kolmogorov-Smirnoff test of both sets of samples returns a $p$-value of $0.88$ for $\alpha$, and $0.49$ for $\beta$, our main large-scale parameter, indicating that the two distributions are compatible. Given this, we therefore use a likelihood of the form
      \begin{equation}
        -2\log p({\bf d}|\vec{\theta})\equiv\chi^2+K=({\bf d}-{\bf t}(\vec{\theta}))^T{\sf C}^{-1}({\bf d}-{\bf t}(\vec{\theta}))+K,
      \end{equation}
      where ${\bf d}$ is the data vector, $\vec{\theta}$ is the set of free parameters in our model, ${\bf t}$ is the theory prediction, ${\sf C}$ is the covariance matrix\footnote{We correct ${\sf C}^{-1}$ by the so-called ``Hartlap factor'' \citep{astro-ph/0608064} to account for the bias in covariance matrices estimated from a finite number of simulations.}, and $K$ is an arbitrary normalisation constant.

      In principle our likelihood depends on 5 free parameters $\{A_{gg},A_{gm},\alpha,\beta,\gamma\}$. However, the theoretical prediction has a linear dependence on the amplitude parameters. Specifically, for a given pair of auto- and cross-corelation measurements
      \begin{equation}
        {\bf t}={\sf T}\cdot{\bf A}\equiv
        \left(
        \begin{array}{cc}
          {\bf t}_{gg} & {\bf 0} \\
          {\bf 0} & {\bf t}_{g\kappa}
        \end{array}
        \right)\cdot
        \left(
        \begin{array}{c}
          A_{gg}\\ A_{gm}
        \end{array}
        \right),
      \end{equation}
      where ${\bf t}_{gg}$ and ${\bf t}_{g\kappa}$ are the theoretical predictions for $A_{gg}=A_{gm}=1$. We assume flat priors on the amplitude parameters, and therefore the log-posterior has a quadratic functional dependence on them. This allows us to marginalise over these parameters exactly. We follow the procedure outlined in \cite{2301.11895}, which further allows us to avoid any volume/projection effects resulting from this marginalisation.

      We assume uniform priors on the slope parameters of the form
      \begin{equation}
        p(\beta)=p(\gamma)=U(-2, 10).
      \end{equation}
      Since the power spectrum turnover has a significantly smoother shape as a function of $\log k$, we sample the logarithm of $\alpha$ instead of $\alpha$ itself, assuming a flat prior of the form
      \begin{equation}
        p(\log_{10}\alpha)=U(-1,1).
      \end{equation}
      This choice of priors has an impact in our determination of the detection significance of the turnover, which we discuss at length in Section \ref{ssec:res.det}.

      To sample the posterior distribution we use the Metropolis-Hastings Markov Chain Monte-Carlo (MCMC) technique as implemented in the {\tt Cobaya} code\footnote{\url{https://cobaya.readthedocs.io/en/latest/}} \citep{2005.05290}. All theoretical calculations were carried out using the Core Cosmology Library\footnote{https://ccl.readthedocs.io/en/latest/}.

    \subsubsection{The projected turnover scale as a standard ruler}\label{sssec:meth.th.cosmo}
      \begin{figure}
        \centering
        \includegraphics[width=0.48\textwidth]{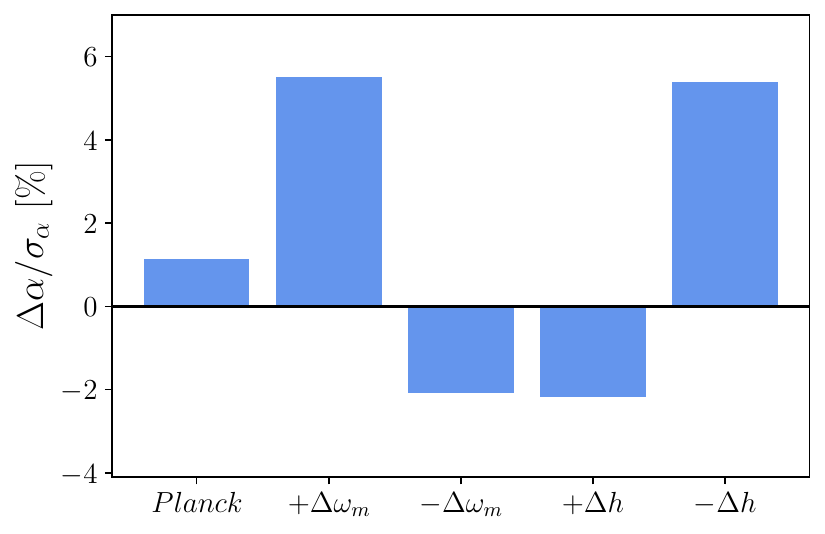}
        \caption{Difference, relative to the statistical uncertainties, between the best-fit value of $\alpha$ and its theoretical standard ruler interpretation as per Eq. \ref{eq:alpha_sruler} for a variety of cosmological models.}\label{fig:bias_alpha}
      \end{figure}
      To interpret our measurement of the power spectrum turnover in a cosmological setting, we must connect the stretch parameter $\alpha$ with the cosmological parameters governing the background expansion of the Universe. As discussed in Section \ref{sssec:meth.th.cl}, a turnover in the matter power spectrum at a comoving scale $k_{\rm TO}$ would map onto a turnover in the angular power spectra at an angular scale $\ell_{\rm TO}\approx k_{\rm TO}d_A(\bar{z})$, where $d_A(\bar{z})$ is the {\sl comoving} angular diameter distance at the median redshift of the sample under study. The model described in Sections \ref{sssec:meth.th.cl} and \ref{sssec:meth.th.model} would predict this turnover at an angular scale $l$ given by $k_{\rm TO}^{\rm fid}\,d_A^{\rm fid}(\bar{z})/\alpha$, where the ``fid'' superscript denotes quantities estimated in the fiducial cosmological model assumed (the best-fit \planck \lcdm parameters, in our case). We can therefore interpret our constraints on $\alpha$ as a measurement of the combination
      \begin{equation}\label{eq:alpha_sruler}
        \alpha=\frac{d^{\rm fid}_A(\bar{z})\,k^{\rm fid}_{\rm TO}}{d_A(\bar{z})\,k_{\rm TO}}.
      \end{equation}

      As described in \cite{1111.2889}, $k_{\rm TO}$ is closely related to the equality scale $k_{\rm eq}$, albeit with a weak dependence on the baryon density $\omega_b\equiv\Omega_bh^2$:
      \begin{equation}\label{eq:keq2kto}
        k_{\rm TO}=\frac{0.194}{\omega_b^{0.321}}k_{\rm eq}^{0.685-0.121\log_{10}(\omega_b)},
      \end{equation}
      with $k_{\rm eq}$ and $k_{\rm TO}$ both in units of $h{\rm Mpc}^{-1}$. Here, the equality scale is defined as $k_{\rm eq}\equiv a_{\rm eq}H(a_{\rm eq})$, where $a_{\rm eq}$ is the scale factor at equality. For a Universe filled with relativistic and non-relativistic species, this is given by
      \begin{equation}
        k_{\rm eq}=\frac{H_*}{c}\sqrt{\frac{2}{\omega_r}}\omega_m,
      \end{equation}
      with $H_*\equiv100\,{\rm km}\,{\rm s}^{-1}\,{\rm Mpc}^{-1})$, and
      \begin{equation}\label{eq:om_r}
        \omega_r\equiv\frac{8\pi G}{3H_*^2}\frac{4\sigma_{\rm SB}T_{\rm CMB}^4}{c^3}(1+0.2271N_{\rm eff})
      \end{equation}
      is the physical density parameter for relativistic species, with $T_{\rm CMB}$ the temperature of the CMB, and $N_{\rm eff}$ the effective number of additional relativistic species. $\sigma_{\rm SB}$ is the Stefan-Boltzmann constant, $G$ is Newton's constant, and $c$ is the speed of light. Fixing $T_{\rm CMB}=2.7255$ \citep{2009ApJ...707..916F}, and $N_{\rm eff}=3.044$ (the value for 3 neutrino species), $\omega_r=4.183\times10^{-5}$, and therefore
      \begin{equation}
        k_{\rm eq}=(0.01043\,{\rm Mpc}^{-1})\,\frac{\omega_m}{0.143}.
      \end{equation}
      Furthermore, we will fix $\omega_b$ to the best-fit value found by \planck, $\omega_b\equiv0.02212$ \citep{1807.06209}. Note that fixing it instead to the value preferred by Big-Bang nucleosynthesis data \citep{2401.15054} leads to negligible differences in $k_{\rm TO}$ of up to $0.35\%$ for reasonable values of $k_{\rm eq}$.

      Combining all the above, we obtain a simplified expression for the dependence of $k_{\rm TO}$ on $\omega_m$ and $h$:
      \begin{equation}
        k_{\rm TO}=\left[0.01114\,{\rm Mpc}^{-1}\right]\,\left(\frac{\omega_m}{0.143}\right)^B\,\left(\frac{h}{0.7}\right)^{1-B},
      \end{equation}
      with $B\equiv0.8853$. The turnover scale is therefore strongly dependent on the physical density of non-relativistic matter, and only weakly on $h$.

      It is not immediately obvious that, for a finite redshift bin, the interpretation of a measured value of $\alpha$ as in Eq. \ref{eq:alpha_sruler} is sufficiently precise. Namely, given a sufficiently wide bin, projection effects from 3D to angular scales can lead to a shift in standard ruler-based observables based on angular statistics \citep{1006.3226}. This is particularly relevant in our case, given the wide range of redshifts covered by the \quaia samples we use. To verify that a cosmological interpretation of $\alpha$ as in Eq. \ref{eq:alpha_sruler} is correct in our case, and quantify any potential biases, we carry out the following exercise:
      \begin{itemize}
        \item We create synthetic data vectors, comprising the same auto- and cross-spectra used in our fiducial analysis, for a set of known cosmologies. Specifically, we consider the best-fit \planck cosmology (which is also the fiducial cosmology used in our analysis), and variations from it in $\omega_m$ and $h$ with $\Delta\omega_m=\pm0.02$ and $\Delta h=\pm0.1$.
        \item We find the best-fit value of $\hat{\alpha}$ for each synthetic data vector, and the corresponding theoretical prediction $\alpha_{\rm th}$ according to Eq. \ref{eq:alpha_sruler} for the true underlying cosmology.
        \item We quantify the difference between the estimated and theoretical strectch parameters $\hat{\alpha}-\alpha_{\rm th}$ relative to our measurement uncertainties.
      \end{itemize}
      The result of this exercise is shown in Fig. \ref{fig:bias_alpha}. We find differences that are significantly smaller than the statistical uncertainties (below 5\%). Similar results are recovered repeating the analysis for each of the individual redshift bins, and for a single redshift bin combining all \quaia sources. We therefore conclude that, at least for the dataset analysed here, it is safe to interpret our measurements of $\alpha$ in terms of the turnover scale as a standard ruler. This calibration study should be repeated, however, for any other dataset, particularly those achieving tighter statistical uncertainties.

\section{Results}\label{sec:res}
  \subsection{Detection of the power spectrum turnover}\label{ssec:res.det}
    \begin{figure*}
      \centering
      \includegraphics[width=0.8\textwidth]{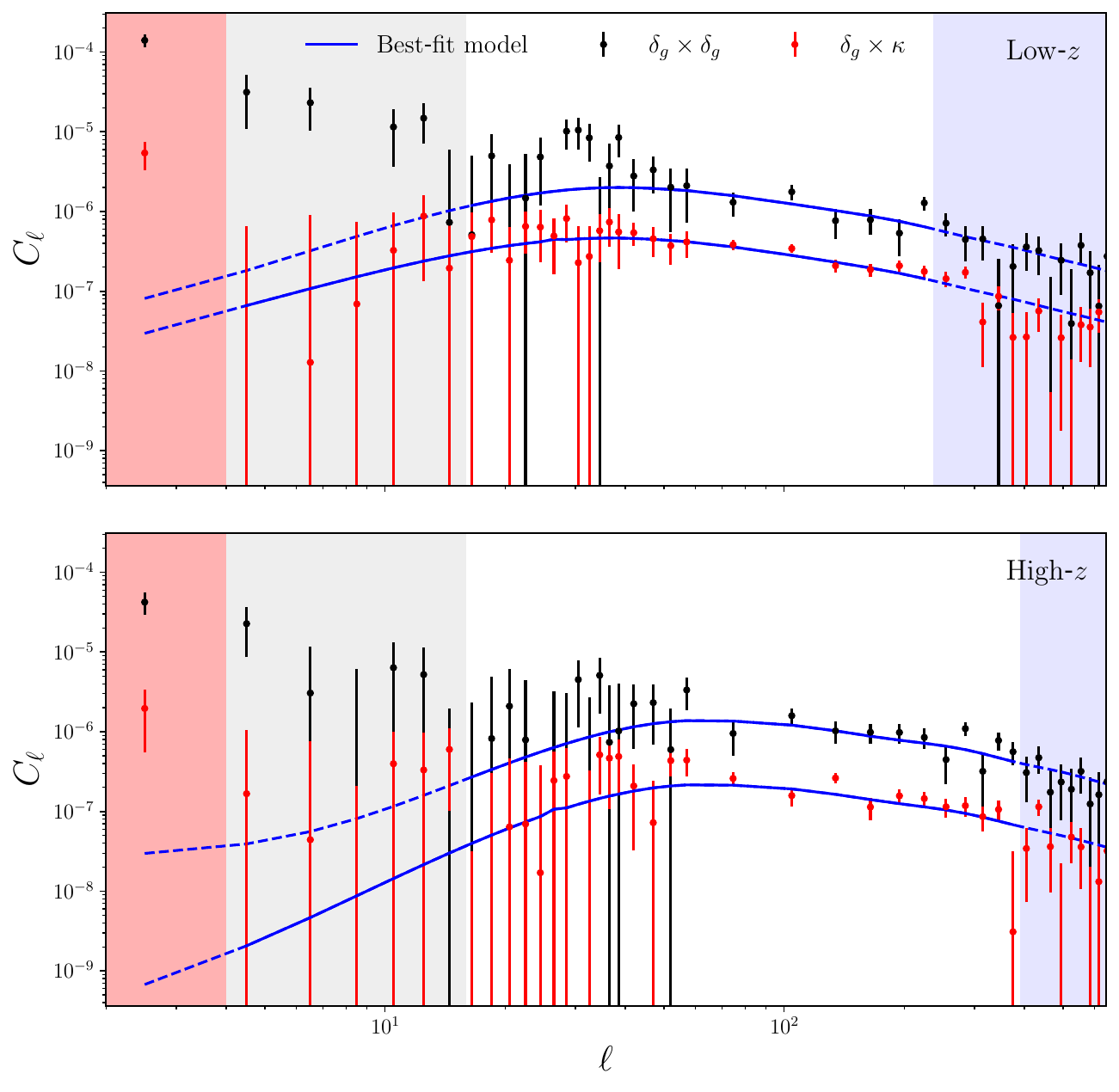}
      \caption{Measured power spectra (quasar auto-spectra in black, quasar-$\kappa$ cross-correlations in red) together with their best-fit model predictions (blue lines). Results are shown for the two Quaia redshift bins (top and bottom panels for low and high redshifts respectively). The red and grey vertical bands show the large scales removed from the fiducial analysis of the auto- and cross-correlations respectively, with the small-scale cuts marked by the blue vertical band.}\label{fig:cl_master}
    \end{figure*}
    \begin{figure}
      \centering
      \includegraphics[width=0.48\textwidth]{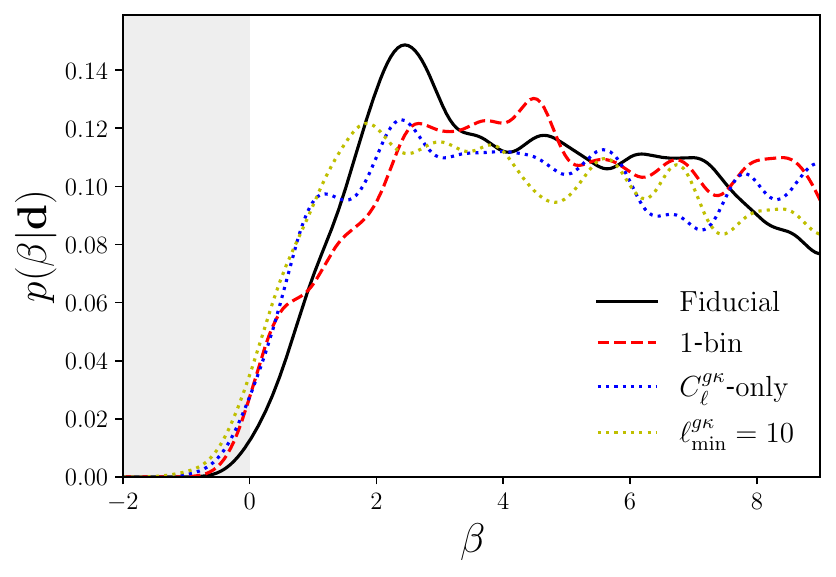}
      \caption{Posterior distribuion of the large-scale slope parameter $\beta$ found in our fiducial analysis (black). Negative values of $\beta$, marked by the grey vertical band, correspond to models with no power spectrum turnover. The dashed red, dash-dotted blue, and dotted yellow lines show the posteriors found when analysing a single redshift bin, discarding the quasar auto-correlation entirely, and increasing the large-scale cut to $\ell_{\rm min}=10$ for the cross-correlation, respectively.}\label{fig:p_beta}
    \end{figure}
    Figure \ref{fig:cl_master} shows our measure power spectra, together with their best-fit predictions according to the model described in Section \ref{sssec:meth.th.model}. The vertical red and grey lines show the large-scales excluded from our fiducial analysis for $C_\ell^{g\kappa}$ and $C_\ell^{gg}$ respectively, with our small-scale cut shown by the vertical blue band. The theoretical prediction extrapolated to these scales is shown as dashed lines. This prediction is a reasonable fit to the data, with a minimum $\chi^2=98.7$ with $N_{\rm dof}=110-7$ degrees of freedom for a probability-to-exceed (PTE) value of ${\rm PTE}=0.60$.    

    Although the theoretical predictions are useful in guiding the eye, it is not immediately evident that the data shows a clear preference for a turnover in the power spectrum, compared with a plateau or a monotonically-decreasing trend. To quantitatively determine our confidence that the turnover is indeed detected, we can use the posterior distribution of $\beta$: since negative values of $\beta$ correspond to spectra with no turnover (with $\beta=0$ a plateau), the detection significance of the turnover can be determined from the probability $p(\beta<0)$. This marginalised posterior is shown in black in Fig \ref{fig:p_beta}. We find that, with our fiducial analysis choices, 99.83\% of the MCMC samples correspond to positive $\beta$ values, corresponding to a $3.1\sigma$ detection of a turnover in the power spectrum. The best-fit value of $\beta$ in our fiducial setup is $\beta=3.22$, steeper than the $\Lambda$CDM value $\beta=1$ (which is nevertheless compatible with our constraints). This result is robust against changes in these choices: repeating our analysis combining all \quaia sources into a single redshift bin (red dashed line), we find $p(\beta<0)=0.0035$ ($2.9\sigma$). Since the quasar auto-correlation is more sensitive to systematic contamination on large scales, it is also interesting to explore the constraints obtained from an analysis involving only the cross-correlations with CMB lensing. This results in a mild reduction of the detection significance $p(\beta<0)=0.0064$ ($2.7\sigma$ -- dash-dotted line in Fig. \ref{fig:p_beta}), but still a clear preference for a turnover. Finally, as a large-scale effect, the detectability of the turnover may depend critically on the lowest multipoles used in the analysis. Since these may be more sensitive to inaccuracies in the covariance matrix or even the shape of the likelihood itself, we repeat the analysis using a more conservative large-scale cut, with $\ell_{\rm min}=10$ for the cross-correlations. The resulting posterior (dotted yellow line) still exhibits a clear preference for the presence of a turnover, with a similar confidence level ($p(\beta<0)=0.009$, or $2.6\sigma$).
    
    We explored the impact of several other analysis choices on this result, including the impact of linear deprojection and its associated transfer function, the impact of the lensing reconstruction transfer function, and the effect of the small-scale cut assumed. Varying these choices did not result in a significant deviation of the result described above: a turnover is preferred at the $\sim2.5$-$3\sigma$ level in all cases explored.

    The methodology used above to quantify the turnover detection significance suffers from a clear flaw. As shown in Fig. \ref{fig:p_beta}, our upper bound on $\beta$ is completely driven by the prior used in our analysis ($\beta<10$). The reason for this is that, due to the size of the measurement uncertainties, we are not able to place a lower bound on the amplitude of the power spectrum at very large scales. Thus, although the data does show a preference for a turnover, the exact slope towards the left of that turnover is difficult to determine. Although we could address this by increasing the upper edge of the prior used, it is not clear that this would solve the problem for any finite value of $\beta_{\rm max}$. Instead, we repeat our analysis now sampling a different parameter, $B$, related to $\beta$ via
    \begin{equation}
      \beta={\rm tan}(\pi B/4).
    \end{equation}
    This transformation maps the infinite interval $\beta\in(-\infty,\infty)$ to the finite one $B\in(-2,2)$, while preserving the interpretation of $B<0$ as models with no turnover, and $B=\beta=1$ as the value in the fiducial \lcdm model. Repeating our analysis, but sampling over $B$ with a flat prior over the range above, we obtain a detection significance of $p(B<0)=0.02$, or $2.3\sigma$. Thus, although we are now able to sample a wider range of values for $\beta$, the detection significance has been reduced (while still remaining compelling). This is easy to understand: this reparametrisation is equivalent to imposing a prior on $\beta$ of the form $p(\beta)\propto(1+\beta^2)^{-1}$ which suppresses the probability of values $|\beta|\gg 1$, and hence reducing the amplitude of the high-$\beta$ tail that caused the lower $p$-values reported above. We conclude, therefore, that while the presence of a turnover in the power spectrum seems to be clearly preferred by the data, the confidence level with which this turnover is detected, depends on the prior assumed on $\beta$.

    For a Bayesian analyst this is the end of the story -- the result depends on the prior and for sensible priors on $\beta$ we get a significant, but not decisive detection. To satisfy the working class, we also carry out a frequentist test, which does not depend on any $\beta$ priors, as an alternative way to quantify this detection significance. We generate 1000 random realisations of the measured power spectra, drawn from a multivariate normal with a mean given by the best-fit theoretical prediction derived from our fiducial analysis, and the covariance matrix of our data. We then find the best-fit values of all free parameters in the model ($\alpha$, $\beta$, $\gamma$, and the power spectrum amplitudes) for each realisation, and count the number of realisations for which the best-fit $\beta$ is negative. Through this exercise, we find $p(\beta<0)=0.02$ (i.e. 20 realisations were found with $\beta<0$), corresponding to a $2.3\sigma$ detection. A similar exercise drawing realisations with a mean given by a theoretical prediction with $\beta=0$ (i.e. a limiting model with no turnover) found only $32$ out of 1000 realisations (i.e. $p=0.032$) with a best-fit value of $\beta$ above that found in our data. In other words, the limiting case of a power spectrum with no clear turnover, is disfavoured at the $2.15\sigma$ level.

    In conclusion, we find that our data shows preference for a turnover in the matter power spectrum at the $2.3$-$3\sigma$ level, although the exact confidence level depends on the methodology used to estimate it.

  \subsection{Measurement of the turnover scale}\label{ssec:res.scale}
    \begin{figure}Measurement
        \centering
        \includegraphics[width=0.5\textwidth]{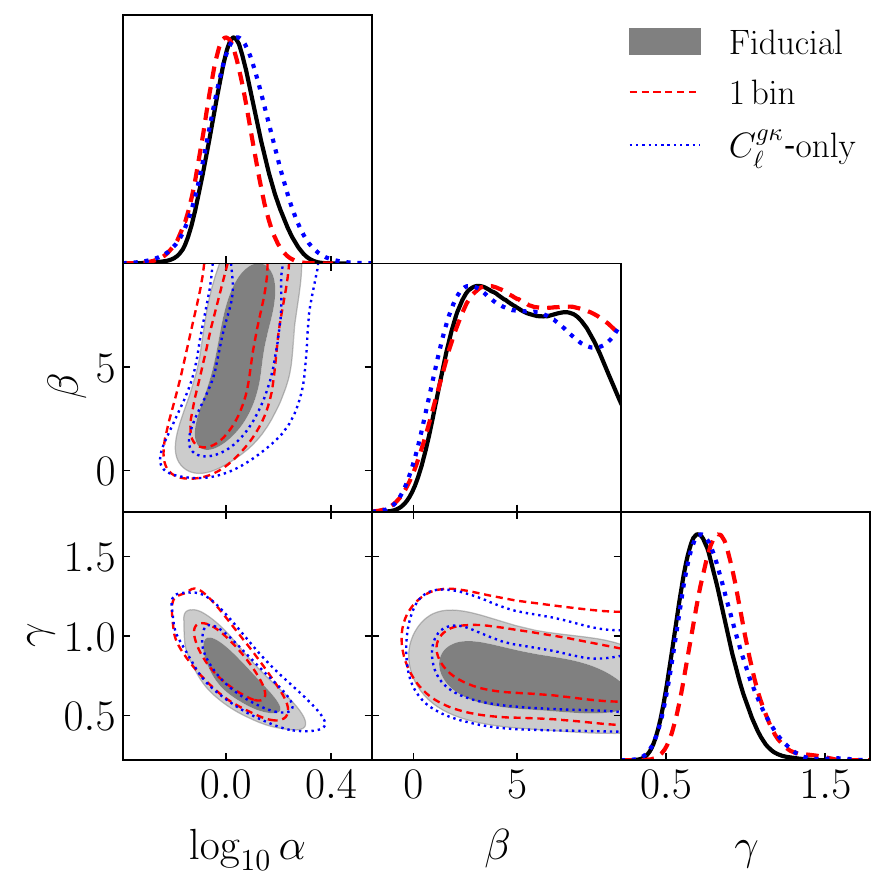}
        \caption{Constraints on the parameters of the model described in Section \ref{sssec:meth.th.model}: the large- and small-scale slopes $\beta$ and $\gamma$, and the stretch parameter $\alpha$. Results are shown for the fiducial combination of $C_\ell^{gg}$ and $C_\ell^{g\kappa}$ (grey) in two redshift bins, for the quasar-CMB lensing cross-correlation alone (dotted blue), and for the analysis of \quaia in a single redshift bin (dashed red).}
        \label{fig:params_triangle}
    \end{figure}
    \begin{table}
      \centering
      \renewcommand{\arraystretch}{1.5}
      \normalsize
      \begin{tabular}{|l|l|}
        \hline
        Analysis settings & $\log_{10}\alpha$\\[0.5ex] 
        \hline
        $1.\,{\rm Fiducial}$ & $0.038^{+0.100}_{-0.091}$\\[0.5ex]
        $2.\,C_\ell^{g\kappa}\,\,{\rm only}$ & $0.054^{+0.122}_{-0.108}$\\[0.5ex]
        $3.\,{\rm Low}$-$z$ & $0.191^{+0.182}_{-0.198}$\\[0.5ex]
        $4.\,{\rm High}$-$z$ & $-0.000^{+0.094}_{-0.085}$\\[0.5ex]
        $5.\,{\rm 1\,\,bin}$ & $0.003^{+0.092}_{-0.083}$\\[0.5ex]
        $6.\,{\rm No}\,\,\kappa\,\,{\rm TF}$ & $0.089^{+0.082}_{-0.107}$\\[0.5ex]
        $7.\,{\rm No\,\,deproj.\,\,TF}$ & $0.013^{+0.103}_{-0.102}$\\[0.5ex]
        $8.\,{\rm No\,\,deprojection}$ & $0.031^{+0.120}_{-0.093}$\\[0.5ex]
        $9.\,\ell_{\rm min}^{g\kappa}=2$ & $0.040^{+0.098}_{-0.092}$\\[0.5ex]
        $10.\,\ell_{\rm min}^{g\kappa}=10$ & $0.033^{+0.105}_{-0.095}$\\[0.5ex]
        $11.\,\ell_{\rm min}^{gg}=30$ & $0.096^{+0.115}_{-0.139}$\\[0.5ex]
        $12.\,\ell_{\rm min}^{gg}=50$ & $0.102^{+0.141}_{-0.142}$\\[0.5ex]
        $13.\,k_{\rm max}=0.1\,{\rm Mpc}^{-1}$ & $-0.032^{+0.088}_{-0.074}$\\[0.5ex]
        $14.\,k_{\rm max}=0.05\,{\rm Mpc}^{-1}$ & $-0.011^{+0.088}_{-0.113}$\\[0.5ex]
        $15.\,{\rm No\,\,wiggles}$ & $0.063^{+0.077}_{-0.077}$\\[0.5ex]
        $16.\,{\rm Fixed}\,\,\gamma$ & $-0.082^{+0.051}_{-0.049}$\\
        $17.\,{\rm Fixed}\,\,\beta$ & $-0.069^{+0.081}_{-0.056}$\\
        \hline
      \end{tabular}
      \caption{Constraints on the logarithm of the stretch parameter $\alpha$ characterising the position of the power spectrum turnover for different analysis choices. These results are also presented in Fig. \ref{fig:lalpha_tests}, and further details are discussed in the main text.}
      \label{tab:lx}
    \end{table}
    \begin{figure*}
      \centering
      \includegraphics[width=0.9\textwidth]{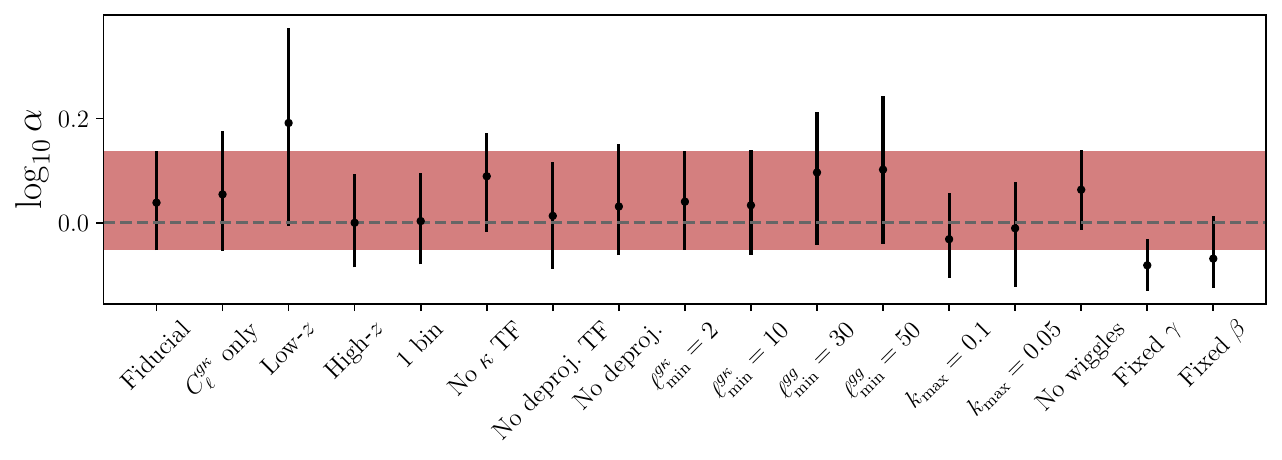}
      \caption{Constraints on the logarithm of the stretch parameter $\alpha$ characterising the position of the power spectrum turnover for different analysis choices. The fiducial constraints are marked by the pink-shaded band, with the value expected in our fiducial $\Lambda$CDM model ($\alpha=1$) marked by the horizontal dashed line. The numerical results displayed here are listed in Table \ref{tab:lx}.}
      \label{fig:lalpha_tests}
    \end{figure*}
    Having established the level of confidence with which a turnover in the power spectrum is present in our data, we now turn to the measurement of the scale at which this turnover takes place. As discussed in Section \ref{sssec:meth.th.model}, this is parametrised in terms of the stretch parameter $\alpha$, with $\alpha=1$ corresponding to a turnover scale compatible with that of the best-fit \planck \lcdm cosmology.

    Fig. \ref{fig:params_triangle} shows, in grey, the constraints on the three model parameters, $\{\alpha,\beta,\gamma\}$, in our fiducial setup. Our 68\% confidence level (C.L.) constraints on $\log_{10}\alpha$ from this fiducial analysis are
    \begin{equation}
      \log_{10}\alpha=0.038^{+0.100}_{-0.091},\hspace{12pt}\alpha=1.093^{+0.283}_{-0.208},
    \end{equation}
    corresponding to a $\sim25\%$ determination of the turnover scale. We carry out a number of tests to verify that this measurement is robust against a large number of potential systematics and analysis choices. The results of these tests are shown in Fig. \ref{fig:lalpha_tests} and summarised in Table \ref{tab:lx}.
    
    Our measurement is robust to potential large-scale systematic contamination in the galaxy overdensity maps. Increasing the smallest multipole over which the galaxy auto-correlations are used, $\ell_{\rm min}^{gg}$ (rows 11 and 12), or removing the galaxy auto-correlation altogether (row 2) does not lead to significant changes in our results. In particular, the latter choice results in a $\sim 20\%$ increase in the uncertainties, confirming that most of the information is concentrated in the $\delta_g\times\kappa$ cross-correlation. These results are shown in blue in Fig. \ref{fig:params_triangle}. As further evidence that contamination in $\delta_g$ does not affect our results significantly, we repeated our analysis omitting the contaminant deprojection stage. The resulting constraints (row 8) are in good agreement with our fiducial results (a shift in $\log_{10}\alpha$ of less than $0.1\sigma$).

    These results are also largely insensitive to the scale dependence of our data on the very largest angular scales, where potential inaccuracies in the model or the covariance matrix could be relevant. Repeating our analysis using a more conservative scale cut $\ell_{\rm min}^{g\kappa}=10$, or a more ambitious $\ell_{\rm min}^{g\kappa}=2$ (rows 9 and 10, respectively), we obtain equivalent constraints, with shifts below $0.1\sigma$ in $\log_{10}\alpha$. Consequently, a potential mis-modelling of the large-scale transfer functions due to deprojection and CMB lensing reconstruction would not affect our results. Omitting these transfer functions (rows 6 and 7) leads to shifts of less than $0.2\sigma$ in the final constraints.

    The stretch parameter $\alpha$ quantifies a potential shift in the scale dependence of the matter power spectrum with respect to our fiducial model. Although the most prominent feature that can in principle drive the constraints on $\alpha$ is the turnover, both the broadband curvature of the power spectrum on scales smaller than this turnover and the BAO wiggles can also act as anchors and effective standard rulers. It is worth noting that this power spectrum curvature has been indeed used in recent works as an effective standard ruler to constrain cosmological parameters, particularly $H_0$ \citep{2008.08084,2112.10749,2204.02984}. Treating the small-scale slope $\gamma$ as a free parameter should minimise our dependence on the power spectrum curvature, and indeed we see in Fig. \ref{fig:params_triangle} that there is a clear degeneracy between $\gamma$ and $\alpha$. Thus it is worth quantifying how much constraining power we lose by suppressing this information. Repeating our analysis fixing $\gamma$ to its fiducial value $\gamma=1$ (row 16) results in constraints on $\log_{10}\alpha$ that are indeed significantly tighter: $\log_{10}\alpha=-0.082^{+0.051}_{-0.049}$, corresponding to a factor $\sim2$ smaller uncertainties, and a difference of $\sim1.6\sigma$ with respect to the fiducial position of the turnover scale ($\alpha=1$). In contrast, fixing the large-scale slope to $\beta=1$ instead (see row 17 in Table \ref{tab:lx}) leads to a milder $\sim25\%$ decrease in the error on $\log_{10}\alpha$, and a less-significant $\sim1\sigma$ downwards shift in its best-fit value (see Fig. \ref{fig:lalpha_tests}). Thus, although both assumptions about the power spectrum slope on scales above and below the turnover play a significant role in determining its position, the impact of the better-measured small-scale slope is larger. The recovered value of $\gamma$ in our fiducial analysis is $\gamma=0.82\pm0.16$, compatible with 1 within $1.1\sigma$. Fixing $\beta=1$ does not change the uncertainty with which $\gamma$ is determined, since constraints on both parameters are driven by different scale ranges, but it pushes its preferred value $\sim0.3\sigma$ closer to $1$: $\gamma=0.87\pm0.16$. Our preferred model is therefore compatible with the \lcdm expectation.

    The $\alpha$-$\gamma$ degeneracy depends on the range of small scales used in the analysis, since these govern the precision with which we can determine $\gamma$. On the other hand, extending the small-scale range may compromise the validity of the linear bias model used in this analysis. We find that our results are reasonably robust with respect to the choice of small-scale cuts. Increasing the scale cut to $k_{\rm max}=0.1\,{\rm Mpc}^{-1}$ (row 10) leads to a $\sim20\%$ decrease in the statistical uncertainties, accompanied by a $\sim0.7\sigma$ downwards shift in $\log_{10}\alpha$. On the other hand, choosing a more conservative cut $k_{\rm max}=0.05\,{\rm Mpc}^{-1}$ produces a $\sim0.4\sigma$ downwards shift and a $\sim10\%$ error increase.

    The statistical uncertainties of our measurements, coupled with the relatively strong projection effects caused by the broad redshift distributions used here, makes it unlikely to recover significant information from the BAO wiggles with \quaia. To verify that indeed our constraints on $\alpha$ are driven by the turnover scale and not the BAO, we repeat our analysis using a linear power spectrum template given by the ``no-wiggle'' parametrisation of \cite{astro-ph/9709112}, which lacks the BAO signature. The resulting constraints on $\alpha$ (row 15) are compatible with our fiducial results, with no increase in the statistical uncertainties due to the lack of a BAO anchor.
    \begin{figure}
      \centering
      \includegraphics[width=0.47\textwidth]{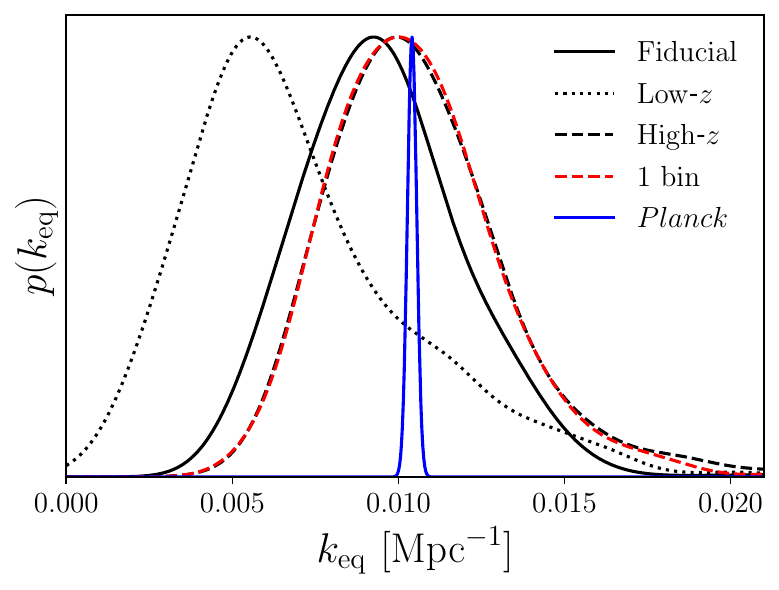}
      \caption{Constraints on the equality scale $k_{\rm eq}$ derived from our measurements of the stretch parameter $\alpha$ fixing all cosmological parameters entering the angular diameter distance to the \planck best-fit values. Results are shown for our fiducial 2-bin analysis (solid black), for each individual redshift bin (dotted and dashed black), and for a single-bin analysis (dashed red), with the \planck constraints shown in blue.}
      \label{fig:keq}
    \end{figure}

    Since the turnover scale is a standard ruler (i.e. a constant comoving scale), the angular scale at which it manifests in the projected statistics explored here scales with the redshift of the sources under study (or, more precisely, with the angular diameter distance to these redshifts). Repeating our analysis for samples at different redshifts is therefore a strong test that our measurements are indeed enabled by a standard ruler. We repeat our analysis including only one of the \quaia redshift bins studied here (rows 3 and 4 for the low- and high-redshift bins, respectively), and for the case in which all \quaia sources are combined into a single redshift bin (row 5). In all cases the constraints are compatible with those found in our fiducial analysis. From the statistical uncertainties of these measurements we can confirm that the majority of the constraining power is concentrated in the higher redshift bin (where the turnover scale appears at larger $\ell$), with the low redshift bin recovering $\sim90\%$ larger errors. This result is not surprising: the turnover scale subtends a smaller angle at higher redshifts, pushing the turnover to larger $\ell$s and providing us with a broader lever arm to detect it and use it as a standard ruler. The results found with a single bin, shown as dashed red lines in Fig. \ref{fig:params_triangle}, are also in good agreement with our fiducial measurements, and recover similar statistical uncertainties.

    Assuming a known cosmology, it is interesting to explore how our constraints on $\alpha$ translate into a measurement of the horizon scale at equality $k_{\rm eq}$. With cosmological parameters fixed to the best-fit \planck values (i.e. our fiducial cosmology), the constraints on $k_{\rm eq}$ found in our fiducial analysis, as well as from each of the two redshift bins separately, and from a single bin combining all \quaia sources, are shown in Fig. \ref{fig:keq}. We obtain a $\sim20\%$ measurement of $k_{\rm eq}$:
    \begin{equation}
      k_{\rm eq}=0.0093^{+0.0025}_{-0.0021}\,{\rm Mpc}^{-1},
    \end{equation}
    in excellent agreement with the value preferred by \planck, $k_{\rm eq}^{\sl Planck}=0.01041\pm0.00014\,{\rm Mpc}^{-1}$. We also find our inferred value of the turnover scale $k_{\rm TO}=0.0157^{+0.0026}_{-0.0021}\,h{\rm Mpc}^{-1}$ to be in good agreement with the measurement of \cite{2302.07484} using the 3D clustering of quasars, $k_{\rm TO}=0.0176^{+0.0019}_{-0.0018}\,\,h{\rm Mpc}^{-1}$.

  \subsection{Cosmological constraints from the turnover scale}\label{ssec:res.cosmo}
    \begin{figure}
      \centering
      \includegraphics[width=0.47\textwidth]{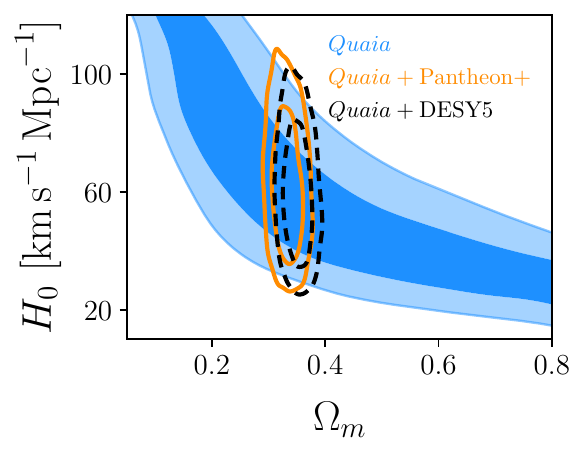}
      \caption{Constraints in the $(\Omega_m,H_0)$ plane derived from our fiducial measurement of the stretch parameter $\alpha$ using \quaia (blue). The solid orange and dashed black contours show the constraints on $H_0$ obtained when combining our measurement with constraints on $\Omega_m$ derived from uncalibrated supernova data from \pantheonp and \desyf, respectively.}
      \label{fig:contours_OmH0}
    \end{figure}
    \begin{figure*}
      \centering
      \includegraphics[width=0.9\textwidth]{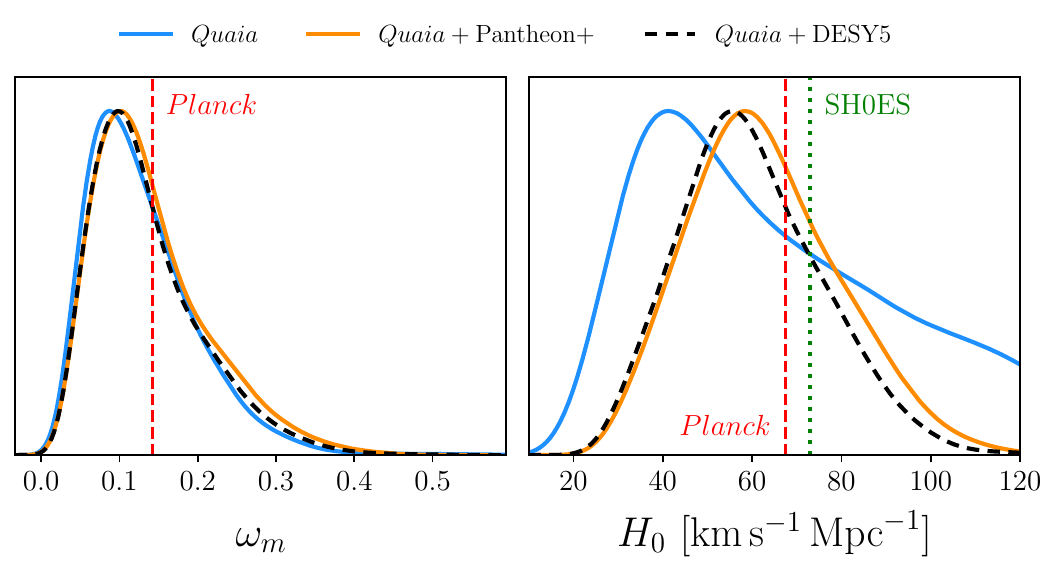}
      \caption{1-dimensional posterior distribution for $\omega_m\equiv\Omega_m h^2$ and $H_0$ obtained from our measurement of the turnover scale alone (blue) and in combination with uncalibrated supernova data from \pantheonp (orange) and \desyf (dashed black). The constraints on $\omega_m$ are virtually independent of any external data.}
      \label{fig:const_om_H0_1d}
    \end{figure*}
    As discussed in Section \ref{sssec:meth.th.cosmo}, assuming a fixed baryon density $\omega_b$, the turnover scale $k_{\rm TO}$ depends exclusively on the matter and radiation densities $\omega_m$ and $\omega_r$. The angular diameter distance $d_A(z)$, connecting $k_{\rm TO}$ with the measured stretch parameter $\alpha$, in turn depends on both $\Omega_m$ and $h$. Assuming the standard model value for $N_{\rm eff}=3.044$, $\omega_r$ can be expressed solely in terms of the CMB temperature, which we will write as $T_{\rm CMB}=\Theta_{\rm CMB}\,T_{\rm CMB}^*$, where $T_{\rm CMB}^*\equiv2.7255$ is the value measured by COBE-FIRAS \citep{2009ApJ...707..916F}. The measurements presented in the previous section therefore depend on three parameters, $\vec{\theta}\equiv\{\omega_m,h,\Theta_{\rm CMB}\}$, and can be used to place constraints on them. To do so, we assume that all information is encoded in the inferred value of $\alpha$ (or, rather, we choose to ignore all other potential information coming from the full shape of the power spectrum, and concentrate only on the turnover scale), and therefore the constraints on $\vec{\theta}$ can be derived from the posterior distribution of $\alpha$:
    \begin{equation}
      p(\vec{\theta}|{\bf d})\propto p(\alpha(\vec{\theta})|{\bf d}).
    \end{equation}
    We calculate the function $p(\alpha|{\bf d})$ using Gaussian kernel density reconstruction from the MCMC chains, and then use the {\tt emcee} package \citep{1202.3665} to sample the resulting distribution.

    Fixing the CMB temperature, we can use our measurements of $\alpha$ to place constraints on a particular combination of $\Omega_m$ and $h$. These constraints are shown in blue in Fig. \ref{fig:contours_OmH0}. Although our measurement of $\alpha$ is not capable of constraining $\Omega_m$ or $H_0$ on their own, we can combine them with current measurements of the distance-redshift relation from uncalibrated\footnote{By this we mean SNe analysed without distance ladder calibrators.} type Ia supernovae (SNe), which constrain $\Omega_m$, to derive a constraint on $H_0$. This is similar to the ``inverse distance ladder'' technique used to derive constraints on $H_0$ from uncalibrated SNe and BAO measurements. The measurement in this case is independent of any assumptions regarding the sound horizon at recombination, depending only on the equality scale instead. The results obtained by combining our measurements with SNe from the \pantheonp \citep{2202.04077} and \desyf \citep{2401.02929} samples are shown in orange and black in Fig. \ref{fig:contours_OmH0}, and correspond to
    \begin{align}\label{eq:h0_const1}
      &H_0=62.7\pm 17.2\,{\rm km}/{\rm s}/{\rm Mpc}\,\,({\sl Quaia}+{\sl Pantheon}),\\\label{eq:h0_const2}
      &H_0=59.8\pm 16.2\,{\rm km}/{\rm s}/{\rm Mpc}\,\,({\sl Quaia}+{\rm DESY5}).
    \end{align}
    Unsurprisingly, given the large uncertainties, our measurements are compatible with the values of $H_0$ measured by both calibrated SNe \citep{2112.04510}, and \planck \citep{1807.06209}. As we discussed earlier, discarding information from the shape of the power spectrum on scales smaller than the turnover has a significant impact on our constraining power. Using the measurement of $\alpha$ obtained fixing the small-scale slope $\gamma$ results in an $H_0$ value, when combined with \desyf, of $H_0=80.1\pm 11.9$. This higher value, related to the low value of $\alpha$ obtained in this case (see Fig. \ref{fig:lalpha_tests}), is nevertheless in reasonable agreement with both SNe and CMB estimates (within $0.6\sigma$ and $1.2\sigma$, respectively).

    Interestingly, and in agreement with the measurements of \cite{2302.07484} using the three-dimensional clustering of eBOSS quasars, we find that our measurement of the turnover scale as a standard ruler is most sensitive to the physical matter density parameter $\omega_m\equiv\Omega_mh^2$. This allows us to obtain constraints on this parameter that are virtually independent from any external data. Fig. \ref{fig:const_om_H0_1d} shows the 1-dimensional posterior constraints found on $H_0$ and $\omega_m$ (right and left panel, respectively). The constraints on $\omega_m$ are driven by the \quaia measurement of the turnover scale. At 68\% confidence level, our fiducial constraints on this parameter are
    \begin{equation}\label{eq:om_const}
      \omega_m= 0.114^{+0.081}_{-0.053}.
    \end{equation}
    This constraint is in good agreement with the value preferred by \planck ($\omega_m=0.142\pm0.001$), and with the measurement of the same parameter made by \cite{2302.07484} from their measurement of the turnover scale ($\omega_m=0.159^{+0.041}_{-0.037}$).

\begin{figure}
      \centering
      \includegraphics[width=0.47\textwidth]{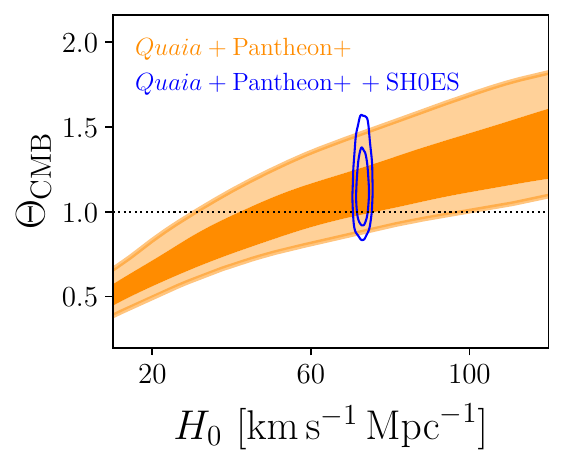}
      \caption{Constraints on the local expasion rate $H_0$ and on the ratio of the CMB temperature to the COBE-FIRAS value $\Theta_{\rm CMB}$. Results are shown for our measurement of the power spectrum turnover in combination with uncalibrated supernova data from \pantheonp (orange), and including an external constraint on $H_0$ from calibrated supernova data by SH0ES (blue).}
      \label{fig:contours_H0tCMB}
    \end{figure}

    Finally, the sensitivity of the equality scale to the energy density of radiation should allow us to place a constraint on the temperature of the CMB. Using our measurement of $\alpha$ in combination with data from uncalibrated SNe allows us to place constraints on a particular combination of $\Theta_{\rm CMB}$ and $H_0$. These are shown in orange in Fig. \ref{fig:contours_H0tCMB}. Combining these constraints with any estimate of $H_0$ then allows us to measure the CMB temperature {\sl independent from measurements of the CMB blackbody spectrum.}
    Using the direct measurement of $H_0$ from low-redshift calibrated SNe by SH0ES \citep{2112.04510} leads to the constraints shown in blue in Fig. \ref{fig:contours_H0tCMB}, corresponding to
    \begin{equation}\label{eq:tcmb_const}
      T_{\rm CMB} = 3.10^{+0.48}_{-0.36}\,{\rm K},
    \end{equation}
    and in good agreement with the blackbody measurement by COBE-FIRAS \citep{2009ApJ...707..916F}. Alternatively, assuming the COBE-FIRAS value for $T_{\rm CMB}$, our measurements can be used to constrain the number of relativistic species in the early Universe, parametrised by $N_{\rm eff}$. From Eq. \ref{eq:om_r}, we can write $\Theta_{\rm CMB}=[1+0.2271\,N_{\rm eff}]^{1/4}$, and we then find
    \begin{equation}
      N_{\rm eff}=3.0^{+5.8}_{-2.9},
    \end{equation}
    compatible with the standard value for 3 neutrino families $N_{\rm eff}=3.044$.

\section{Conclusions}\label{sec:conc}
  The turnover caused by the different evolution of density perturbations during the radiation-dominated and matter-dominated epochs is the most prominent feature of the power spectrum. The relatively large scales at which it appears (approximately the size of the horizon at the matter-radiation equality -- $k_{\rm eq}\sim0.01\,{\rm Mpc}^{-1}$) has, however, made its direct detection in the clustering of galaxies particularly challenging.
  
  In this paper we have used data from the \quaia quasar sample, covering one of the largest volumes probed by a large-scale strucure survey, in combination with CMB lensing data from \planck, to detect this turnover, and measure the scale at which it takes place. This represents one of only a handful of studies aimed at making this measurement, and the first one to employ the cross-correlation between galaxies and CMB lensing to do so. Since this cross-correlation is significantly less affected by large-scale systematic contamination in the galaxy sample than the standard 3D auto-correlations used in the past, this measurement represents a strong test of the \lcdm model.

  We determine the turnover and its scale using only the scale dependence of the measured angular power spectra, marginalising over their amplitudes. The shape of the power spectrum is modelled in terms of three parameters: the spectral slope above and below the turnover, and the scale at which the turnover takes place. The evidence for the presence of a turnover is then quantified in terms of the posterior probability for a positive large-scale power spectrum slope (the negative small-scale slope being significantly easier to determine from the data). Indeed, we find that the data shows a clear preference for positive large-scale slopes, and the associated evidence for a turnover runs from 2.3 to 3.1$\sigma$, depending on the method used to quantify it (using both Bayesian and frequentist approaches). The detection is driven by the cross-correlation measurement, and is robust to multiple analysis choices, including scale cuts and redshift binning, as well as large-scale systematics.

  These measurements translate into a $\sim20\%$ determination of the comoving scale of the turnover. The measurement is driven by the turnover itself, and not from the curvature of the power spectrum on smaller scales, used in previous studies to obtain improved cosmological constraints. Assuming that the small-scale dependence of the power spectrum is known leads to a factor of $\sim2$ reduction in the error on the turnover scale. Assuming the cosmological parameters preferred by \planck data, the inferred equality scale $k_{\rm eq}=0.0093^{+0.0025}_{-0.0021}\,{\rm Mpc}^{-1}$ is in reasonable agreement with the value measured from CMB data. The measurement of the turnover scale is highly robust to the potential presence of large- and small-scale systematics, the mis-modelling of quasar bias, and to various analysis choices (see Table \ref{tab:lx}). Furthermore, we have shown that, as a standard ruler, our measurements are completely driven by the power spectrum turnover, and receive no significant information from the BAO wiggles.

  As we have shown, the measured value of the stretch parameter $\alpha$ can be interpreted as an estimate of the combination $d_A(\bar{z})k_{\rm TO}$, where $d_A(\bar{z})$ is the angular diameter distance to the median redshift of the sample and $k_{\rm TO}$ is the turnover scale (related to $k_{\rm eq}$ via Eq. \ref{eq:keq2kto}). This is a good approximation even for the broad redshift bins used in this analysis. The combination $d_A(\bar{z})k_{\rm TO}$ depends on the physical matter density parameter $\omega_m$, the radiation density parameter (given in terms of the CMB temperature $T_{\rm CMB}$ for a fixed number of relic species), and the expansion rate $H_0$. Fixing the CMB temperature to the value measured by COBE-FIRAS, we have shown that, in fact, our measurements are mostly dependent on $\omega_m$, and can be used to constrain this parameter in the absence of any other data (see Eq. \ref{eq:om_const}). Using uncalibrated supernova data to constrain $\Omega_m$ (i.e. using an inverse distance ladder approach) we can then place a constraint on $H_0$ which, although not competitive with current constraints from supernova, CMB, or even BAO-based inverse distance ladder techniques, is independent of the sound horizon at recombination and decoupling (see Eqs. \ref{eq:h0_const1} and \ref{eq:h0_const2}). Finaly, using calibrated and uncalibrated supernovae to constrain both $\omega_m$ and $H_0$, we can use our measurement to measure the CMB temperature independently of any primary CMB data or, alternatively, the number of relativistic species in the early Universe $N_{\rm eff}$. The resulting measurement of $T_{\rm CMB}$ is in good agreement with the COBE-FIRAS value (Eq. \ref{eq:tcmb_const}), and the constraints on $N_{\rm eff}$ are consistent with the standard value for three neutrino families.

  Future datasets will be able to significantly improve upon current measurements of the power spectrum turnover, and thus exploit it to improve and validate the cosmological constraints obtained through other probes. This will be achieved by covering larger volume and reducing the shot noise of the galaxy samples used. Ongoing spectroscopic experiments, such as DESI \citep{2022AJ....164..207D}, will provide larger samples of high-redshift quasars, and other high-redshift targets, such as Lyman-break galaxies could be ideally suited for this purpose. Future experiments, such as the Maunakea Spectroscopic Experiments \citep{2019arXiv190404907T}, or MegaMapper \citep{1907.11171}, will significantly increase the volume probed by pushing to higher redshifts, and obtain denser samples of of spectroscopic sources, thus increasing the precision of this measuring by a large factor \citep{2302.07484}. The constraining power of the turnover as a cosmological observable will also benefit from measurements at different redshifts, in order to break the degeneracy between $\Omega_m$ and $H_0$ for a single redshift. As we have shown, the turnover scale can also be measured effectively using projected clustering statistics from photometric surveys, as long as large-scale observational systematics can be kept under control, and a reliable estimate of the sample redshift distribution can be achieved. In particular, the Rubin Observatory's Legacy Survey of Space and Time \citep[LSST,][]{0805.2366}, or the {\sl Euclid} satellite \citep{1110.3193}, will be a ideal datasets for this analysis. Spectro-photometric experiments, such as SPHEREx \citep{2404.11017}, covering very large volumes with multiple tracers, will also be optimally suited for these studies. Other LSS probes able to cover ultra-large scales, such as 21cm intensity mapping, may also be able to exploit the turnover \citep{1305.6928,1507.03550,2202.13828}, assuming foreground removal effects can be sufficiently calibrated. In all these cases, as we have shown, cross-correlations with CMB lensing will be vital in order to obtain improved measurements of the turnover scale, and validate them against the impact of large-scale systematic contamination. Future CMB datasets, such as the Simons Observatory \citep{1808.07445}, or CMB-S4 \citep{1610.02743}, will produce considerably more sensitive maps of the CMB lensing convergence that will enable significantly more precise studies of the power spectrum turnover.

\section*{Acknowledgements}
  We thank Pedro Ferreira, Carlos Garc\'ia-Garc\'ia, Maria Vincenzi, Sebastian von-Hausseger, and the anonymous referee for useful comments and discussions. DA acknowledges support from UKSA under grant ST/Y005902/1 and from the Beecroft Trust. GF acknowledges the support of the  European Union’s Horizon 2020 research and innovation program (Grant agreement No. 851274). We made extensive use of computational resources at the University of Oxford Department of Physics, funded by the John Fell Oxford University Press Research Fund.

\bibliography{main}

\end{document}